\documentclass{article}

\usepackage{arxiv}
\usepackage[utf8]{inputenc} 
\usepackage[T1]{fontenc}    
\usepackage{hyperref}       
\usepackage{url}            
\usepackage{booktabs}       
\usepackage{amsfonts}       
\usepackage{nicefrac}       
\usepackage{microtype}      
\usepackage{lipsum}
\usepackage[T1]{fontenc}
\usepackage{authblk}
\usepackage{graphicx}
\usepackage{float}
\usepackage{amsmath}
\usepackage{array}
\usepackage{multirow}
\usepackage{diagbox}
\usepackage{algorithm}
\usepackage{algorithmicx}
\usepackage{algpseudocode}
\usepackage[labelfont=bf]{caption}

\usepackage{apacite}
\usepackage{natbib}
\usepackage{xcolor}
\usepackage{bm}
\usepackage{makecell}
\usepackage[misc]{ifsym}
\usepackage[toc,page]{appendix}
\usepackage[labelsep=space]{caption}
\hypersetup{linktocpage}
\usepackage{epstopdf}

\AtBeginDocument{

}

\title{High-frequency financial market simulation and flash crash scenarios analysis: an agent-based modelling approach}
\author[1,2*]{Kang Gao}
\author[2]{Perukrishnen Vytelingum}
\author[1]{Stephen Weston}
\author[1]{Wayne Luk}
\author[1]{Ce Guo}
\affil[1*]{Department of Computing, Imperial College London, London SW7 2AZ, UK}
\affil[2]{Simudyne Limited, London EC3V 9DS, UK}
\affil[*]{Corresponding author(s). E-mail(s): kang.gao18@imperial.ac.uk}

\begin{document}
\maketitle

\begin{abstract}
\hspace{0.3cm} This paper describes simulations and analysis of flash crash scenarios in an agent-based modelling framework. We design, implement, and assess a novel high-frequency agent-based financial market simulator that generates realistic millisecond-level financial price time series for the E-Mini S\&P 500 futures market. Specifically, a microstructure model of a single security traded on a central limit order book is provided, where different types of traders follow different behavioural rules. The model is calibrated using the machine learning surrogate modelling approach. Statistical test and moment coverage ratio results show that the model has excellent capability of reproducing realistic stylised facts in financial markets. By introducing an institutional trader that mimics the real-world Sell Algorithm\footnote{See \cite{securities2010findings} for detailed description.} on May 6th, 2010, the proposed high-frequency agent-based financial market simulator is used to simulate the Flash Crash that took place that day. We scrutinise the market dynamics during the simulated flash crash and show that the simulated dynamics are consistent with what happened in historical flash crash scenarios. With the help of Monte Carlo simulations, we discover functional relationships between the amplitude of the simulated 2010 Flash Crash and three conditions: the percentage of volume of the Sell Algorithm, the market maker inventory limit, and the trading frequency of fundamental traders. Similar analyses are carried out for mini flash crash events. An innovative "Spiking Trader" is introduced to the model, aiming at precipitating mini flash crash events. We analyse the market dynamics during the course of a typical simulated mini flash crash event and study the conditions affecting its characteristics. The proposed model can be used for testing resiliency and robustness of trading algorithms and providing advice for policymakers.

\end{abstract}

\keywords{Agent-based Model \and Financial Market Simulator \and High-frequency Data \and Flash Crash}

\textbf{JEL Classification} C52 $\cdot$ C63 $\cdot$ G17

\section*{Acknowledgements}
\hspace{0.3cm} The support of the UK EPSRC (Grant Nos. EP/L016796/1, EP/N031768/1, EP/P010040/1,  EP/S030069/1 and EP/V028251/1), Xilinx, and Intel is gratefully acknowledged. Kang Gao holds a China Scholarship Council-Imperial Scholarship.

\section*{Statements and Declarations}
\textbf{Competing Interests} The authors declare that they have no further conflict of interest.

\newpage

\section{Introduction} \label{sectionintroduction}
\hspace{0.3cm} With the advent of electronic financial markets for the exchange of securities, the electronic centralized limit order book has become the standard market mechanism for transaction matching and price discovery. This form of order book offers market participants a more liquid market system with a small bid-ask price spread, increased market depth and decreased transaction times.

\hspace{0.3cm} Algorithmic trading is commonly defined as the use of computer algorithms to automatically make trading decisions, submit orders, and carry out post-submission order management. In the past decade algorithmic trading has grown rapidly across the world and has become the dominant way securities are traded in financial markets, currently generating more than half of the volume of U.S. equity markets. Constantly improving computer technology and its application by both traders and exchanges, together with the evolution of market micro-structure, automation of price quotation and trade execution have together enabled faster trading. Nowadays the speed of order submission has become a principal characteristic for distinguishing trading agents. Market participants known as high-frequency traders are capable of trading hundreds of times in a second, using fast algorithms and specialized network connections with exchanges. high-frequency traders are often orders of magnitude faster in order submission than other traders, and even other trading algorithms.

\hspace{0.3cm} The rise of algorithmic trading and high-frequency trading has had broad impacts on financial markets, especially on the price discovery process and market price stability. One conspicuous impact is the increasingly frequent "flash crash" in major financial markets. The flash crashes comprise large and rapid changes in the price of an asset that does not coincide with changes in economic fundamental value for the asset. The flash crash events have occurred in markets that are among the largest and most liquid exchanges in the world. One representative flash crash event is the famous 2010 Flash Crash, which happened in the U.S. stock market on May 6th, 2010. During this flash crash event, one market participant's algorithm caused a sharp price drop in the E-mini S\&P futures market. The flash crash soon spread to other futures markets and equity markets. The market price fell almost 6\% in just several minutes, while the bulk of losses was recovered nearly as quickly. The 2010 Flash Crash led to turmoil market conditions and caused huge market value loss. As for the cause of the 2010 Flash Crash, \cite{kirilenko2017flash} show that the key events in the 2010 Flash Crash have clear relationships with regard to algorithmic trading.

\hspace{0.3cm} The 2010 Flash Crash seems to be singular because of the fact that no following events have rivalled its depth, breadth, and speed of price movement. Nevertheless, flash crashes on a smaller scale happen frequently. These events are termed mini flash crashes (\citealt{johnson2012financial}). According to \cite{johnson2012financial}, there were more than 18,000 mini flash crashes that are identified in the U.S. equity market between 2006 and 2011. As scaled-down versions of the 2010 Flash Crash, mini flash crashes are abrupt and severe price changes occur in an extremely short time period (\citealt{golub2012high}). Mini flash crashes are attracting great research interest because their frequent occurrence could destabilize the financial market and undermine investor confidence (\citealt{golub2012high}).

\hspace{0.3cm} The flash crash episodes\footnote{We use "flash crashes", "flash crash episodes" and "flash crash scenarios" interchangeably in this article.}, including large flash crash and mini flash crash events, are of significant concern to researchers, practitioners, and policymakers. Financial markets in which price changes are orderly and reflect proper changes in valuation factors are desirable. However, flash crash episodes could potentially disorganise such a desirable market and cause adverse consequences for financial stability if they were to impede investment by undermining investor confidence in the price at which securities could be transacted (\citealt{karvik2018deeds}). To prevent flash crash episodes from becoming more frequent and longer-lasting, it is important to understand how such episodes arise. In this paper we explore the dynamics during both large flash crashes and mini flash crashes. The main methodology employed here is financial market simulation in agent-based models. The agent-based financial market simulation provides realistic synthetic financial market data and a testbed for exploring dynamics during flash crash episodes and conditions that influence the characteristics of flash crash episodes.

\hspace{0.3cm} Financial market simulation based on agent-based models is a promising tool for understanding the dynamics of financial markets. With huge potential academic and industrial value, agent-based financial market simulation has gained extensive research attention in recent years. Financial market simulation by agent-based models is an exciting new field for exploring behaviours of financial markets. An agent-based financial market simulation consists of a number of distinct agents that follow predetermined rules in a manner analogous to how real-world trader behaves in reality. Unlike traditional economic theories, there is no equilibrium assumption in agent-based financial markets. In addition, traders are no longer assumed to have rational behaviours as in traditional economic theories. The removal of these assumptions makes agent-based financial market simulation more realistic than traditional equilibrium-based economic and financial theories. These advantages of agent-based financial market simulation make it possible to explore complex phenomena such as flash crash episodes in modern financial markets, which is unachievable with traditional equilibrium-based theories. 

\hspace{0.3cm} Various agent-based simulators have been developed in the literature. However, there are still gaps in creating ideal agent-based financial market simulators that are capable of generating synthetic high-frequency market data that are realistic. Specifically, most existing agent-based financial markets are of lower frequency such as daily or weekly. To explore market dynamics that involve high-frequency trading, a higher simulation frequency is needed. In addition, instead of using full exchange protocols, many simulators make assumptions about the price formation process and use mathematical formulas to approximate the matching engine. This significantly undermines the realism of the simulator. Last but not the least, the proper calibration and validation of agent-based financial market simulation are still an open problem.  

\hspace{0.3cm} To sum up, there are two \textbf{challenges} that this paper aims to address:
\begin{itemize}
\item C1: To design and implement a high-frequency agent-based financial market simulator with full exchange protocols, and with proper calibration and validation process to reproduce a realistic artificial financial market.
\item C2: Under the proposed agent-based financial market simulator, investigate the market dynamics during flash crashes (including both large flash crashes and mini flash crashes) and explore the conditions that influence the characteristics of flash crashes.
\end{itemize}

\hspace{0.3cm} Motivated by the above challenges, we developed a novel high-frequency financial market simulator to narrow the existing gaps. The simulator is then employed to explore the dynamics during flash crashes and the conditions that affect flash crashes. Broadly speaking, our \textbf{contributions} in this paper are three-fold:
\begin{itemize}
\item A high-frequency agent-based financial market simulator is implemented, with each simulation step corresponding to 100 milliseconds. Full exchange protocols (limit order books) are implemented to simulate the order matching process. In this way, we provide a microstructure model of a single security traded on a central limit order book in which market participants follow fixed behavioural rules. The model is calibrated using the machine learning surrogate modelling approach. As for model validation, statistical test and moment coverage ratio results show that the simulation is capable of reproducing realistic stylised facts in financial markets.
\item Under the framework of the proposed high-frequency agent-based financial market simulator, the 2010 Flash Crash is realistically simulated by introducing an institutional trader that mimics the real-world Sell Algorithm on May 6th, 2010. We investigate the market dynamics during the simulated flash crash and show that the simulated dynamics are consistent with what happened in historical flash crash scenarios. We then explore the conditions that could have influenced the characteristics of the 2010 Flash Crash. According to our Monte Carlo simulation, three conditions significantly affect the amplitude of the 2010 Flash Crash: the percentage of volume (POV) of the Sell Algorithm, market maker inventory limit, and the trading frequency of fundamental traders. In particular, we found that the relationship between the amplitude of the simulated 2010 Flash Crash and the POV of the Sell Algorithm is not monotonous, and so is the relationship between the amplitude and the market maker inventory limit. For the trading frequency of fundamental traders, the higher the frequency, the smaller the amplitude of the simulated 2010 Flash Crash.
\item Similar analysis is carried out for mini flash crash events. An innovative type of trader called "Spiking Trader" is introduced to the agent-based financial market simulator, creating more price shock and precipitating more mini flash crash events. Market dynamics for a typical simulated mini flash crash event are analysed. We also explore the conditions that could influence the characteristics of mini flash crash events. Experimental results show that the market maker inventory limit significantly affects both the frequency and amplitude of mini flash crash events. However, the trading frequency of fundamental traders shows no obvious influence on mini flash crash events in our experiments.
\end{itemize}


\hspace{0.3cm} The \textbf{novelty} of our approach lies in several features. Firstly, the proposed agent-based financial market simulator has a higher frequency than most other simulators in the literature. Our simulation step is at milliseconds level, which allows for the investigation of high-frequency dynamics in the simulated financial market, while most simulation models in literature adopt larger simulation steps of 1 second or 1 minute. Secondly, we explore the influence of different market configurations on the amplitude of the 2010 Flash Crash. To the best of our knowledge, there are few similar experiments in the existing literature. Thirdly, an innovative type of trader named "Spiking Trader" is proposed to precipitate more mini flash crash events. Fourthly, the experiments that explore the conditions that influence the frequency and amplitude of mini flash crash events are also a novelty of this work.

\hspace{0.3cm} The remainder of the article is organized as follows. Section~\ref{background} presents general background on the agent-based financial market simulation and an overview of previous research about flash crash events.  Section~\ref{modelstructure} shows the structure and details for the proposed agent-based model, while Section~\ref{modelcalibration} presents the model calibration process and model validation results. Section~\ref{flashcrashscenarios} and Section~\ref{miniflashcrashscenarios} provide simulation and analysis for the 2010 Flash Crash scenario and mini flash crash scenarios, respectively, in the framework of the proposed agent-based financial market simulation. Section~\ref{conclusion} concludes and gives directions for future work.

\section{Background and Related Work} \label{background}

\subsection{Agent-based Financial Market simulation}
\hspace{0.3cm} An agent-based model (ABM) is a computational simulation
driven by the individual decisions of programmed agents (\citealt{todd2016agent}). ABMs are often used in simulating financial markets. In agent-based simulated financial markets, an agent's objective is to "digest the large amounts of time series information generated during a market simulation, and convert this into trading decisions" (\citealt{lebaron2001builder}). With the advantage of capturing the heterogeneity of agents and diversity of the underlying economic system, ABMs provide a promising alternative to traditional equilibrium-based economic models. 

\hspace{0.3cm} \cite{gode1993allocative} build an agent-based model with only zero-intelligence traders to simulate financial markets. Those zero-intelligence traders are not able to think strategically, or do any advanced learning, or statistical modelling of the financial market. Surprisingly, results show that zero-intelligence traders can trade very effectively in the simulated market. The prices tend to converge to the standard equilibrium price and market efficiency tends to reach a very high level. According to their experimental results, they argue that some stylised facts in financial markets may rely more on institutional design rather than actual agent behaviour. Agent-based models are also proposed to model the "Trend" and "Value" effects in financial markets. Chiarella designed an agent-based model composed of two types of traders: fundamentalists and chartists (\citealt{chiarella1992dynamics}). With only two types of traders, lots of dynamic regimes that are compatible with empirical evidence can be generated in the simulated artificial financial market. An extension of the Chiarella model is proposed in \cite{MAJEWSKI2020103791}. The extended model adds a new type of trader called noise trader and allows the fundamental asset value to have a long term drift. The extended Chiarella model is capable of reproducing more realistic price dynamics. This extended Chiarella model in \cite{MAJEWSKI2020103791} forms the basis of the proposed agent-based financial market simulation in this paper. A more complex agent-based model for financial market simulation is proposed in \cite{mcgroarty2019high}. Five different types of traders are present in the simulated market: market makers, liquidity consumers, momentum traders, mean reversion traders, and noise traders. Their model is capable of replicating most of the existing stylised facts of limit order books, such as autocorrelation of returns, volatility clustering, concave price impact, long memory in order flow, and the presence of extreme price events. Those stylised facts have been observed across different asset classes and exchanges in real financial markets. The successful replication of these stylised facts indicates the validity of their agent-based simulation model. It is shown that agent-based financial market simulation is capable of generating artificial financial markets with realistic macro behaviours.

\hspace{0.3cm} The prevalence of electronic order books and automated trading permanently change the way the market works. It is virtually impossible to infer meaningful relationships between market participants using traditional mathematical methods because of the complexity of electronic financial markets. Instead, agent-based financial market simulation has been gradually getting popularity in the market microstructure literature. An agent-based simulated financial market offers an experimental environment for examining market features and characteristics. It also provides plenty of artificial financial market data for analysis. \cite{hayes2014agent} develop an agent-based model for use by researchers, which offers the capability of capturing the organization of exchanges, the heterogeneity of market participants, and the intricacies of the trading process. Agent-based models can also provide regulators with an experimental environment that helps to comprehend complex system outcomes. In other words, it allows for a clearer examination of the relationship between micro-level behaviour and macro outcomes. For example, \cite{darley2007nasdaq} test the regulatory changes that came with decimalization in the NASDAQ market using agent-based financial market simulation. The agent-based models of the NASDAQ market shed light on how these changes would impact market function.

\hspace{0.3cm} To summarise, agent-based financial market simulation simplifies complex financial system simulation by including a set of individual agents, a topology and an environment. Different agent-based models in the literature focus on different practical problems in financial markets. In this paper, we focus on agent-based models applied to flash crash analysis. In the following section, we provide a literature review of flash crash episodes.

\subsection{Flash Crash Episodes}
\hspace{0.3cm} During the 2010 Flash Crash, over a trillion dollars were wiped off the value of US equity markets in an event that has been largely attributed to the rapid rise of algorithmic trading and high-frequency trading \cite{kirilenko2017flash}. The base indices in both the futures and securities market experienced a rapid price fall of more than 5\% in just several minutes, after which the bulk of the price drop was recovered nearly as fast as it fell. The staff from CFTC and SEC present a thorough report on what happened during the 2010 Flash Crash event (\citealt{securities2010findings}). They identified an automated execution algorithm that sold a large number of contracts as the main catalyst for the flash crash. The Sell Algorithm, which was activated on the E-mini S\&P 500 futures market, kept pace with the market aiming at selling around 9\% of the previous minute's trading volume (\citealt{securities2010findings}). Even though no negative impact was known previously, this process triggered a cascade of panic selling by market participants that employ high-speed automated trading systems. The consequent "hot-potato" effect, where those market participants rapidly acquired and then liquidated positions among themselves, resulted in rapid and extreme price decline.

\hspace{0.3cm} Flash crash episodes have attracted attention after the 2010 Flash Crash event. Several months after the crash, the staff from regulatory authorities released a report that highlighted the important role of a large seller in initiating the flash crash event (\citealt{securities2010findings}). It is reported that although high-frequency traders appear to have exacerbated the magnitude of the crash, they do not actually trigger the flash crash. Though high-frequency traders played a role in creating the so-called "hot-potato" effect, the flash crash would very likely have been avoided without the overly simplistic sell algorithm based on volume alone. There is also a lot of academic research on flash crash episodes. For example, \cite{kirilenko2017flash} applied purely empirical approaches to understanding the causes of the 2010 Flash Crash. They use regression analysis on a unique dataset that is labelled with the identities of all market participants. It is demonstrated that in responding to the activity of the Sell Algorithm, high-frequency traders caused the "hot-potato" effect that exacerbated the price drop. This is consistent with the \cite{securities2010findings} report. \cite{paddrik2012agent} develop an agent-based model of the E-mini S\&P futures applied to flash crash analysis. A general flash crash in price is replicated in their model. However, they only reproduce a rough shape of the flash crash price behaviour, and detailed analyses of trader behaviours and market depth are absent. \cite{karvik2018deeds} developed an agent-based model to analyse the flash crash episodes in the sterling-dollar forex market. They emphasize the important role of high-frequency traders in the emergence of flash crash episodes. The proposed approach in this paper is partly inspired by their work. \cite{paddrik2017effects} explore how the levels of information can be used to predict the occurrence of flash crash events. Their findings suggest that some stability indicators derived from limit order book information are capable of signalling a high likelihood of an imminent flash crash event. 

\hspace{0.3cm} There are also different angles of view for flash crash episodes in literature. \cite{paulin2019understanding} design and implement a hybrid microscopic and macroscopic agent-based approach to investigate the conditions that give rise to the "electronic contagion"\footnote{Electronic contagion refers to the contagion phenomenon in financial markets that results from interactions between trading algorithms, instead of human traders.} of flash crash events. Their results demonstrate that the flash crash contagion between different assets is dependent on portfolio diversification, behaviours of algorithmic traders, and network topology. It is also stressed that regulatory interventions are important during the propagation of flash crash distress. \cite{menkveld2019flash} look at the flash crash event from the perspective of cross-arbitrage. They find that the breakdown of cross-arbitrage activities between related markets plays an important role in exacerbating the flash crash event. \cite{kyle2020large} analyse price impact during flash crash events in their market microstructure invariance model. It is shown that the actual price declines in flash crash events are larger than the predicted price impact. \cite{madhavan2012exchange} argues that the flash crash episodes are linked directly to the current market structure, mostly the pattern of volume and market fragmentation. He further suggests that a lack of liquidity is the critical issue that requires the greatest policy attention to prevent future flash crash events. Similarly, \cite{borkovec2010liquidity}  explicitly owe the flash crash in ETFs to an extreme deterioration in liquidity. Their results are consistent with the liquidity provision behaviour in financial markets. \cite{golub2012high} analyse mini flash crashes, which are the scaled-down versions of the 2010 Flash Crash. It is shown that mini flash crashes also have an adverse impact on market liquidity and are associated with fleeting liquidity phenomenon.

\hspace{0.3cm} The above provides various analyses for the occurrence of flash crash episodes. However, despite extensive work on analysing the flash crash episodes, the exact causes of the flash crash episodes are still not clear. In this paper, we investigate and analyse the flash crash episodes through the lens of agent-based financial market simulation. In this sense, our work is similar to the work in \cite{karvik2018deeds} and \cite{paddrik2012agent}. Nevertheless, we offer a much more extensive and detailed analysis of the simulated flash crash event which, to the best of our knowledge, is the most fine-grained analysis in current literature. Specifically, we realistically simulate the 2010 Flash Crash event in our simulation, and divide the simulated flash crash event to several phases. For each phase, detailed analyses about traders behaviours and market dynamics are presented. To the best of our knowledge, this fine-grained simulation and analysis is not reported before in literature. By dividing the whole flash crash event into different phases and examine trader behaviours and market dynamics for each phase, we shed light on the cause for flash crash events. In addition, controlled experiments under different model settings and trader behaviours are carried out in the developed agent-based simulation framework, which provide insights about how to prevent the happening of detrimental flash crash events. The basis of our agent-based model is the extended Chiarella model in \cite{MAJEWSKI2020103791}, which comprises fundamental traders, momentum traders, and noise traders. We further divide momentum traders into long-term momentum traders and short-term momentum traders, and introduce market makers to the model. The motivation for introducing these types of traders and their interactions will be presented in section~\ref{modelsetup}. It is shown that the proposed model is capable of generating realistic artificial financial time series. Within the framework of the proposed realistic agent-based financial market simulation, special types of agents are introduced to trigger flash crash episodes in the simulated financial market. In this way, simulated flash crash episodes are scrutinized and analysed.



\section{Model Structure} \label{modelstructure}
\hspace{0.3cm} This section presents the set-up and components of the proposed agent-based high-frequency financial market simulator. 

\subsection{Model Set-up} \label{modelsetup}
\hspace{0.3cm} We denote the price of a stock at time $t$ as $P_t$. The price in this work refers to mid-price, which is the mean value of the best bid price and the best ask price at time $t$:
\[
P_t := 0.5 * b^0_t + 0.5 * a^0_t
\]
where $b^0_t$ and $a^0_t$ denote best bid and best ask price at time $t$, respectively. The price dynamics are exactly the same as the price dynamics in real stock exchanges - a limit order book is built to accept all orders and trades are generated by the match engine. All the orders submitted by various traders in one simulation period, including limit orders and market orders, constitute the aggregated virtual "demand" and "supply" in that period. The aggregated virtual "demand" and "supply" depend on the trading strategies of various types of market participants. The market participants are assumed to be heterogeneous in their trading decisions. According to the Extended Chiarella model (\citealt{MAJEWSKI2020103791}), a popular set of traders in an artificial financial market includes momentum traders, fundamental traders, and noise traders. This composition of traders is capable of capturing the trend and value effects in financial markets. We further extend the model by dividing momentum traders into two groups: short-term momentum traders and long-term momentum traders. In this way, heterogeneity is introduced not only between groups of traders, but also inside a specific group of traders. In addition, in an artificial financial market with full exchange protocols, market maker is an indispensable component to create realistic limit order book behaviours.  To sum up, five types of traders are included in our model: fundamental traders, short-term momentum traders, long-term momentum traders, noise traders, and market makers. Each type of trader is associated with several parameters which control the trading behaviours. Trading heuristics, specific parameters and the parameter calibration process will be presented in subsequent sections. 

\subsection{Common Trader Behaviours}
\hspace{0.3cm} Traders in our simulation model have some behaviours in common, for example submitting and cancelling orders. We can assume that there is a certain type of agent called "Base Agent", and all traders in our model inherit the functionalities of this base agent. Specifically, all traders in the model have common behaviours as follows. 

\begin{itemize}
\item Each trader has a parameter $\theta$, which controls the probability of submitting a limit order. The value of $\theta$ depends on the type of traders and also varies during simulation, but the behaviour given a $\theta$ value is common for all types of traders. That is, for each simulation step, if $p \in U(0, 1)\footnote{$p \in U(0, 1)$ returns a value that follows uniform distribution between 0 and 1.} < \theta$, the trader will place an limit order, otherwise no action is taken. Note that the specific side of the order (buy or sell) depends on the trader type and market conditions. The value of $\theta$ could be zero for some traders, which means that the corresponding traders do not submit limit orders. 
\item Each trader has a parameter $\mu$. The function of $\mu$ is identical to parameter $\theta$, except that $\mu$ controls the probability of sending a market order. Similar to $\theta$, the value of $\mu$ is different for different types of traders and varies for different timestamps. For each simulation step, if $p \in U(0, 1) < \mu$, the trader will place a market order, otherwise no action is taken. The specific side of the market order also depends on the trader type and market conditions. Similar to $\theta$, the value of $\mu$ can also be zero for some particular type of traders. 
\item The side for all limit orders and market orders are determined by corresponding trader types and market conditions. A market order is submitted directly to the exchange after the side is determined. For a limit order, the price information is required. Given the order side, a buy limit order will have a price lower than the market mid-price at the corresponding timestamp, while a sell limit order will have a price higher than the market mid-price. The distance between the limit order price and the market mid-price follows a particular distribution, whose type is determined by the trader types. Specifically, the price distance for limit orders from market makers is sampled from a uniform distribution with parameters $0$ and $p^{MM}_{edge}$, and the price distance for limit orders from other types of traders is sampled from a common log-normal distribution with parameters $\mu_{\ell}$ and $\Sigma_{\ell}$. The parameters of these distributions are calibrated to historical price time series stylised facts. The price for a limit order is calculated after the price distance is sampled from the corresponding distribution. 
\item Each type of trader has a parameter $\delta$. Regardless of the order price and side, each limit order has a cancellation probability of $\delta$ at each simulation step. The value of $\delta$ is dependent on the type of trader who places the order, and is calibrated using historical market data to create realistic limit order book behaviours. In the presented model, all traders share the same value for $\delta$ except market makers, who have a higher value of $\delta$. This reflects the fact that orders submitted by market makers tend to have a much higher cancellation / replacement rate.
\item In the proposed model, short-term momentum traders, long-term momentum traders and noise traders all  have a parameter called $\rho$, which controls the ratio between the number of market orders and limit orders placed by the same trader. That is, for each trader of these three types, there is a fixed relationship between $\theta$ and $\mu$: $\mu = \theta * \rho $. This relationship enables a realistic ratio between the number of market orders and limit orders in the simulated financial market and the value of $\rho$ is selected according to historical orders data.
\item The volume $V$ for each order is 100, regardless of limit order or market order.
\end{itemize}

\hspace{0.3cm} Despite the common trader behaviours, each type of trader follows different trading heuristics and has different values for the associated parameters. For example, fundamental traders only submit market orders. Market makers submit limit orders in normal trading time. Only after the inventory limit is hit will market makers submit market orders to reduce their inventory risk. The remaining types of traders submit both limit orders and market orders during trading hours, aiming to maintain a fixed ratio between the number of limit orders and the number of market orders. The remainder of the section describes the trading behaviour heuristics for each type of trader in more detail. Descriptions for all the parameters involved in the proposed model are summarised in Appendix~\ref{modelparameterdescription}.


\subsection{Fundamental Trader (FT)}
\hspace{0.3cm} Fundamental traders make their trading decisions based on the perceived fundamental value of the stock. The fundamental value is denoted as $V_t$. A fundamentalist tends to buy a stock if the stock is under-priced($V_t - P_t > 0$), otherwise he will sell the stock. Following the convention in \cite{MAJEWSKI2020103791}, in this work we assume the aggregated demand of fundamental traders is polynomial to the level of mispricing. Specifically, the aggregated demand of fundamentalists is:

\begin{equation} \label{fundamentalequation}
D_{FT} = \kappa_{1}(V_t - P_t) + \kappa_{2}(V_t - P_t)^3
\end{equation}

where $\kappa_1$ and $\kappa_2 $ controls the overall demand generated by fundamental traders. Since we have $N_{FT}$ fundamental traders, each fundamental trader will contribute $\frac{\kappa_{1}|V_t - P_t| + \kappa_{2}|V_t - P_t|^3}{N_{FT}}$ to the aggregated demand. Since the fundamental traders only submit market orders, the value of $\theta$ for each fundamental trader is set to 0, while the value of $\mu$ for each fundamental trader is set to $\frac{\kappa_{1}|V_t - P_t| + \kappa_{2}|V_t - P_t|^3}{N_{FT}}$. The fundamental value $V_t$ is an exogenous signal that is input to the model. Each fundamental trader is also associated with a parameter $S^{FT}_{interval}$. A fundamental trader only attempts to submit a market order once in every $S^{FT}_{interval}$ simulation steps. A detailed description of the fundamental trader's logic is given in Algorithm~\ref{fundamentallogic}.

\floatname{algorithm}{Algorithm}
\begin{algorithm}
  \caption{Fundamental Trader Logic}
  \label{fundamentallogic}
  \begin{algorithmic}[1]
  \For{ \textbf{each} simulation step}
  \State $\theta \gets  0$
  \State $\mu \gets \frac{\kappa_{1}|V_t - P_t| + \kappa_{2}|V_t - P_t|^3}{N_{FT}}$  
  
  \If { ($simulation\_steps \mod S^{FT}_{interval} == 0$) and ($p \in U(0, 1) < \mu$)}  
  \If {$V_t - P_t > 0$} Place a buy market order
  \ElsIf{$V_t - P_t < 0$} Place a sell market order
  \EndIf
  \Else {\hspace{0.25cm} No market order submitted}
  \EndIf
  
  \EndFor
  \end{algorithmic}
\end{algorithm}

\subsection{Momentum Trader (MT)}
\hspace{0.3cm} Momentum traders are also called "Chartists". This group of traders buy and sell financial assets after being influenced by recent price trends. The assumption is to take advantage of an upward or downward trend in the stock prices until the trend starts to fade. Instead of looking at the fundamental value of the stock, momentum traders focus more on recent price action and price movement. If the stock price has been recently rising, a long position is established; otherwise, momentum traders will enter a short position. In the proposed model, momentum traders can submit both limit orders and market orders. For momentum traders, the ratio between the number of limit orders and market orders is fixed, which is denoted as $\rho$.

\hspace{0.3cm} There are lots of methods to estimate the momentum of stock prices. A common trend signal is the exponentially weighted moving average of past returns with decay rate $\alpha$. This trend signal is denoted by $M_t$:
\begin{equation} \label{momentumequation}
M_{t} = (1 - \alpha)M_{t-1} + \alpha(p_t - p_{t-1})
\end{equation}
where $\alpha$ is the decay rate. Given the trend signal $M_t$, the demand function of momentum traders is denoted as $f(M_t)$. The demand function $f(M_t)$ must satisfy two conditions:
\begin{itemize}
\item $f(M_t)$ is increasing.
\item $f''(M_t) * M_t < 0$
\end{itemize}
where the first condition is consistent with the nature of momentum trading and the second condition imposes the risk-averse assumption to momentum traders. Consistent with \cite{MAJEWSKI2020103791}, here we choose $f(M_t) = \beta \tanh(\gamma M_t)$ with the requirement that $\gamma > 0$. $\gamma$ represents the saturation of momentum traders' demand when momentum signals are very large. This phenomenon is partly due to for example budget constraints and risk aversion, which are prevalent in real momentum traders. $\beta$ controls the overall demand generated by momentum traders. $\beta$ is also assumed to be positive, i.e. the demand of momentum traders is positive when the momentum signal ($M_t$) is positive, otherwise the demand is negative. The choice of this demand function for momentum traders strictly satisfies the two requirements.

\hspace{0.3cm} According to the value of $\alpha$, we further divide the group of momentum traders into two sub-groups: long-term momentum traders (small $\alpha$) and short-term momentum traders (large $\alpha$). The two types of momentum traders also have different values for $\beta$; however, they share the same $\gamma$ value. The proposed model includes $N_{LMT}$ long-term momentum traders and $N_{SMT}$ short-term momentum traders.
\subsubsection{Long-term Momentum Trader (LMT)}
\hspace{0.3cm} Long-term momentum traders are associated with a small value for $\alpha$. According to equation~\ref{momentumequation}, a small $\alpha$ corresponds to slow changes in the momentum signal. Consequently, the momentum signal is smooth and reflects the trend on a longer time scale. In our intra-day simulation, we choose an $\alpha$ value of 0.01 for long-term momentum traders. This would approximately correspond to an intra-day hourly trend. The group of long-term momentum traders will focus on the relatively long-time price trend and calculate their demand function. For limit orders, the virtual aggregated demand from long-term momentum traders is $\beta_L \tanh(\gamma M_t^L)$, while for market orders the virtual aggregated demand from long-term momentum traders is $\rho \beta_L \tanh(\gamma M_t^L)$. As there are $N_{LMT}$ long-term momentum traders, the $\theta$ and $\mu$ for each trader are corresponding quantities divided by $N_{LMT}$. Trading decisions are made according to the common order submission rules. A full description of long-term momentum traders' logic is shown in Algorithm~\ref{longtermmomentumlogic}.

\begin{algorithm}
  \caption{Long-term Momentum Trader Logic}
  \label{longtermmomentumlogic}
  \begin{algorithmic}[1]
  \For{ \textbf{each} simulation step}
  \For{ \textbf{each} outstanding limit orders submitted by this trader}
  \If { $p \in U(0, 1) < \delta$} Cancel this limit order
  \EndIf
  \EndFor
  \State $M_{t} = (1 - \alpha_L)M_{t-1} + \alpha_L(p_t - p_{t-1})$
  \State $\theta \gets \frac{\beta_L \tanh(\gamma M_t)}{N_{LMT}}$
  \State $\mu \gets \theta * \rho$  
  \If { $p \in U(0, 1) < \theta$}  
  \If {$M_t > 0$} Place a buy limit order
  \ElsIf{$M_t < 0$} Place a sell limit order
  \EndIf
  \Else {\hspace{0.25cm} No limit order submitted}
  \EndIf
  
  \If { $p \in U(0, 1) < \mu$}  
  \If {$M_t > 0$} Place a buy market order
  \ElsIf{$M_t < 0$} Place a sell market order
  \EndIf
  \Else {\hspace{0.25cm} No market order submitted}
  \EndIf
  
  \EndFor
  \end{algorithmic}
\end{algorithm}

\subsubsection{Short-term Momentum Trader (SMT)}
\hspace{0.3cm} Compared to the long-term momentum traders, short-term momentum traders are associated with a much larger $\alpha$. Typically, $\alpha$ for short-term momentum trader would be a value close to 1. Following equation~\ref{momentumequation}, the momentum signal is updated very fast when $\alpha$ has a large value. In this circumstance, $M_t$ represents the trend in an extremely short time scale, typical only several ticks in our intra-day simulation. The group of short-term momentum traders aim at exploiting the price trend in a very short time scale, which mimics the behaviour of some real-world high-frequency traders. In our agent-based financial market simulation, we assign a value of 0.9 to $\alpha$ of short-term momentum traders.\footnote{We also run simulation experiments with $\alpha$ having value 0.95 and 0.99. Similar experimental results are obtained in these settings, showing that minor changes in $\alpha$ value will not significantly affect the results.} The trading heuristics for short-term momentum traders are exactly the same as that for long-term momentum traders, except that $\alpha_L$ and $\beta_L$ are replaced by $\alpha_S$ and $\beta_S$, respectively. A full description of short-term momentum traders' logic is shown in Algorithm~\ref{shorttermmomentumlogic}.

\begin{algorithm}
  \caption{Short-term Momentum Trader Logic}
  \label{shorttermmomentumlogic}
  \begin{algorithmic}[1]
  \For{ \textbf{each} simulation step}
  \For{ \textbf{each} outstanding limit orders submitted by this trader}
  \If { $p \in U(0, 1) < \delta$} Cancel this limit order
  \EndIf
  \EndFor
  \State $M_{t} = (1 - \alpha_S)M_{t-1} + \alpha_S(p_t - p_{t-1})$
  \State $\theta \gets \frac{\beta_S \tanh(\gamma M_t)}{N_{smt}}$
  \State $\mu \gets \theta * \rho$  
  \If { $p \in U(0, 1) < \theta$}  
  \If {$M_t > 0$} Place a buy limit order
  \ElsIf{$M_t < 0$} Place a sell limit order
  \EndIf
  \Else {\hspace{0.25cm} No limit order submitted}
  \EndIf
  \If { $p \in U(0, 1) < \mu$}  
  \If {$M_t > 0$} Place a buy market order
  \ElsIf{$M_t < 0$} Place a sell market order
  \EndIf
  \Else {\hspace{0.25cm} No market order submitted}
  \EndIf
  
  \EndFor
  \end{algorithmic}
\end{algorithm}

\begin{algorithm}[H]
  \caption{Noise Trader Logic}
  \label{noiselogic}
  \begin{algorithmic}[1]
  \State $\theta \gets \frac{\sigma_{NT}}{N_{NT}}$
  \State $\mu \gets \theta * \rho$  
  \For{ \textbf{each} simulation step}
  \For{ \textbf{each} outstanding limit orders submitted by this trader}
  \If { $p \in U(0, 1) < \delta$} Cancel this limit order
  \EndIf
  \EndFor
  \If { $p \in U(0, 1) < \theta$}  
  \If {$p \in U(0, 1) < 0.5$} Place a buy limit order
  \ElsIf{$p \in U(0, 1) > 0.5$} Place a sell limit order
  \EndIf
  \Else {\hspace{0.25cm} No limit order submitted}
  \EndIf
  
  \If { $p \in U(0, 1) < \mu$}  
  \If {$p \in U(0, 1) < 0.5$} Place a buy market order
  \ElsIf{$p \in U(0, 1) > 0.5$} Place a sell market order
  \EndIf
  \Else {\hspace{0.25cm} No market order submitted}
  \EndIf
  
  \EndFor
  \end{algorithmic}
\end{algorithm}

\subsection{Noise Trader (NT)}
\hspace{0.3cm} Another group of market participants are noise traders. They are designed so as to capture other market activities that are not reflected by trend-following and value investing. As a result, the cumulative demand of noise traders can be described by a random walk, with each step having an equal probability of placing buy/sell orders. Parameter $\sigma_{NT}$ controls the aggregated demand level from noise traders. There are $N_{NT}$ noise traders in the model, with each noise trader having value $\frac{\sigma_{NT}}{N_{NT}}$ for $\theta$. Similar to momentum traders, noise traders are also associated with a parameter $\rho$, which controls the ratio between the number of limit and market orders. Note that unlike other traders whose $\theta$ and $\mu$ vary according to simulated market conditions, $\theta$ and $\mu$ for noise traders are determined prior to simulation and remain fixed during the simulation process. Algorithm~\ref{noiselogic} gives a description of noise traders' logic in the simulated financial market.

\subsection{Market Maker (MM)}
\hspace{0.3cm} The market makers are another group of traders in the model. The introduction of market makers in the proposed model is aimed at creating realistic limit order book dynamics. Market makers in the proposed model are more complex than previous traders. During normal trading time, market makers only submit quotes to the market. A quote includes one buy limit order and one sell limit order. The simplification that market makers only submit limit orders is motivated by \cite{menkveld2013high}, which finds that around 80\% of market makers' orders are passive. The price of the sell (buy) order is calculated by adding (subtracting) a distance from the mid-price at the corresponding timestamp, where the distance is sampled from a uniform distribution. (The price is rounded to the closest multiple of tick size before being submitted to the exchange.) In alignment with the market making behaviours during the 2010 Flash Crash event (\citealt{securities2010findings}), market makers in the proposed model are associated with a position limit. Specifically, once the inventory of a market maker reaches the position limit, the market maker will stop all active quoting and actively submit market orders to reduce the inventory level. This will continue until the inventory reduces to a certain safe level, which is also a parameter of the model. At this stage, the market maker suspends trading for a certain time period. This resembles the real-world scenarios that market makers tend to suspend trading to check their own trading systems and observe market conditions after some unusual scenarios happen. After this time period, the market maker will restart the normal trading heuristics. Table~\ref{MMorders} presents the corresponding order types that market makers will submit in different trading conditions.

\begin{table}[H]
\centering
\caption{Order types submitted by market makers under different trading conditions}
\begin{tabular}{ | m{2.5cm}<\centering | m{3.5cm}<\centering| m{4cm}<\centering | m{3.5cm}<\centering |} 
\hline
Trading condition & Normal trading & Stressed trading (after inventory limit reached) & Trading suspension \\
\hline
Order type & Limit order  & Market order & None\\ 
\hline
\end{tabular}
\label{MMorders}
\end{table}

\hspace{0.3cm} Since market makers only submit limit orders during the normal trading time, the $\mu$ for each market maker is set to 0. Similar to the case in previous traders, $\delta$ and $\theta$ control the probability of order cancellation and limit order  submission respectively, whose values are calibrated to historical price time series. Note that one difference is that $\theta$ controls the probability of submitting a quote (buy and sell limit orders) for market makers. In addition to $\delta$, $\theta$ and $\mu$, market makers are associated with several extra parameters that correspond to the trading behaviours presented above. The trading edge of market makers is sampled from a uniform distribution, which controls the spread of the quotes submitted by market makers. Specifically, the uniform distribution from which the price distance is sampled has a minimum value of 0 and a maximum value of $p^{MM}_{edge}$. $\varepsilon^{MM}_{limit}$ represents the position limit for each market maker, while $\varepsilon^{MM}_{safe}$ denotes the safe position level. That is, market makers will actively reduce inventory once $\varepsilon^{MM}_{limit}$ is reached, until the position is reduced to $\varepsilon^{MM}_{safe}$ level. $\varepsilon^{MM}_{rest}$ represent the time length for the trading suspension. Algorithm~\ref{MMlogic} presents a full description of market makers' trading heuristics in the proposed model.

\begin{algorithm}
  \caption{Market Maker Logic}
  \label{MMlogic}
  \begin{algorithmic}[1]
  \State Before simulation starts:
      \State \hspace{0.5cm} $\theta \gets \theta_{MM}$
      \State \hspace{0.5cm} $\mu \gets 0$
      \State \hspace{0.5cm} $\delta \gets \delta_{MM}$
      \State \hspace{0.5cm} $Flag \gets 0$
      \State \hspace{0.5cm} $Restart\_step \gets 0$ \\
  \For{ \textbf{each} simulation step}
  \State Get current simulation step $current\_step$
  \If{$|position| >= \varepsilon^{MM}_{limit}$}\\ 
      \hspace{1cm}$Flag \gets 1$ \\
      \hspace{1cm}$Restart\_step = Current\_step + \varepsilon^{MM}_{rest}$
  \EndIf \\
  \If{$(Flag == 1) \land (|position| <= \varepsilon^{MM}_{safe})$} $Flag \gets 0$
  \EndIf \\
  \If{$Flag == 1$}\\
      \hspace{1cm} Cancel all active limit orders
      \If{|position| < 0}\\
      \hspace{1.5cm} Submit a buy market order 
      \Else{\\ \hspace{1.5cm} Submit a sell market order}
      \EndIf \\
  \ElsIf{$Current\_step > Restart\_step$}
      \If { $p \in U(0, 1) < \delta$} \\
      \hspace{1.5cm} Cancel existing quotes by this market maker
      \EndIf \\
      \If { $p \in U(0, 1) < \theta$} \\
      \hspace{1.5cm} Submit a quote(one limit buy order and one limit sell order, with sampled trading edge.)          
      \Else {\\ \hspace{1.5cm} No limit order submitted}
      \EndIf \\
  \EndIf
  \EndFor
  \end{algorithmic}
\end{algorithm}

\subsection{Simulation Dynamics}
\hspace{0.3cm} The above are the five types of agents included in the proposed model. A fully functioning limit order book was implemented, as is the case in most electronic financial markets. The simulation is run in pseudo-continuous time. Specifically, each simulation step represents 100 milliseconds of trading time. Each trading day is divided into $T = 324,000$ steps, corresponding to 9 hours of trading (8:00 to 17:00). The minimum time for execution of transactions is 100 milliseconds, showing that the simulated financial market represents a high-frequency trading environment. We remind here that the fundamental value $V_t$, which is extracted from the historical price time series, is an exogenous signal that is input to the model. 

\hspace{0.3cm} The whole simulation runs as follows. For each step, each trader collects and processes market information. Internal variables associated with each trader are calculated. According to agent type and values of internal variables, actions are taken by the traders. These actions include limit order submission, market order submission, and order cancellation. The programmed matching engine matches these orders and updates the state of the limit order book. Finally, transactions and limit order book status are published to all traders. The whole simulation procedure is shown in Algorithm~\ref{simulationlogic}.

\begin{algorithm}
  \caption{Simulation Procedure}
  \label{simulationlogic}
  \begin{algorithmic}[1]
  \State Simulation Parameter Initialization
  \While {timestamp < market close}
  \State timestamp += {\tiny$ \Delta$}$t$ \\
  \For{ \textbf{each} trader}
  \State Collect latest market information and update internal states
  \If {fundamental trader} 
  \State Fundamental trader logic(Algorithm~\ref{fundamentallogic})
  \EndIf
  \If {long-term momentum trader} 
  \State Long-term momentum trader logic(Algorithm~\ref{longtermmomentumlogic})
  \EndIf
  \If {short-term momentum trader} 
  \State Short-term momentum trader logic(Algorithm~\ref{shorttermmomentumlogic})
  \EndIf
  \If {noise trader} 
  \State Noise trader logic(Algorithm~\ref{noiselogic})
  \EndIf
  \If {market maker}
  \State Market maker logic(Algorithm~\ref{MMlogic})
  \EndIf
  \EndFor \\
  \For{ \textbf{each} limit order, market order, or order cancellation)}
  \State Update limit order book and publish transactions
  \EndFor \\
  \State Publish latest limit order book status
  
  \EndWhile
  \end{algorithmic}
\end{algorithm}

\hspace{0.3cm} We suggest that the proposed five types of traders reflect a sufficiently realistic and diverse market environment. According to \cite{ohara1995book}, there are three major market-microstructure trader types: uninformed traders, informed traders and market makers. The noise traders in our model correspond to uninformed traders, while market makers in the proposed model obviously correspond to the market makers in literature. The remaining three types of traders represent informed traders in our model. Specifically, fundamental traders utilise exogenous information implied by the fundamental value, while the two types of momentum traders exploit the endogenous technical indicator information. In addition, among the informed traders some perceived trading opportunities are based only on an analysis of short-horizon returns, while others focus on market information revealed by long-term return horizons. This is reflected by the division of momentum traders into long-term and short-term momentum traders. Overall, a sea of different informed and uninformed traders in the proposed model compete with each other, with market makers providing liquidity and ensuring realistic limit order book behaviours. In conclusion, the proposed model with five types of traders represents a complete range of micro-behaviours of real financial markets.

\subsubsection{Fundamental Value from Kalman Smoother}
\hspace{0.3cm} The only remaining unknown variable is the fundamental value of the stock. The simulation can proceed only if the fundamental value is known and is exogenously input to the model. One difficulty is the non-observability of the fundamental value. According to the economic literature, the fundamental value of a stock equals the expected value of discounted dividends that the company will pay to the shareholders in the future. However, this methodology requires extremely strong assumptions about the future dynamics of the stock dividends. Furthermore, this approach can never reflect the intra-day change of fundamental value, while the consensus fundamental value can indeed vary during the trading day due to the continuous feed of events and news. 

\hspace{0.3cm} In this paper, we propose a new method which is to apply Kalman Smoother (\citealt{1556088}) directly to the stock price time series to get the hidden fundamental value. In accordance with \cite{MAJEWSKI2020103791}, we assume the fundamental value $V_t$ is a hidden variable of a linear dynamical system. The observations are the actual prices traded in real markets. The specific Kalman Smoothing algorithm used here can be found in \cite{byron2004derivation}. The algorithm is applied to the price time series for each trading day to extract the corresponding fundamental value time series for that day.\footnote{The algorithm is already implemented in Python package "pykalman".}

\section{Model Calibration and Validation} \label{modelcalibration}
\hspace{0.3cm} In this section we present the methodology for calibrating the agent-based financial market. Calibration means finding an optimal set of model parameters to make the model generate the most realistic simulated financial market. Firstly we describe the real data and the associated stylised facts in financial markets. Next we define the distance between historical and simulated stylised facts, which acts as the loss function in the calibration process. The parameters calibration workflow is presented, followed by detailed validation of the proposed high-frequency financial market simulator.

\subsection{Calibration Target: Data and Stylised Facts for Realistic Simulation} \label{data}
\hspace{0.3cm} In the model calibration process, real financial market data is essential for setting up the calibration target. We collected high-frequency limit order book data of E-mini S\&P 500 futures, from May 3rd, 2010 to May 6th, 2010\footnote{Evening trading sessions are excluded. That is, the data span from 8:00 in the morning to 5:00 in the afternoon for each day.}. We select the most liquid contract as the calibration target, which is the contract expires in June 2010. Our dataset comprises high-frequency information for 10 levels of limit order book update, both the buy side and sell side. 

\hspace{0.3cm} Financial price time series data display some interesting statistical characteristics that are commonly called stylised facts. According to \cite{sewell2011characterization}, stylized facts refer to empirical findings that are so consistent (for example, across a wide range of financial instruments and different time periods) that they are accepted as truth. A stylized fact is a simplified presentation of an empirical finding in financial markets. A successful and realistic financial market simulation is capable of reproducing various stylised facts. These stylised facts include fat-tailed distribution of returns, autocorrelation of returns, and volatility clustering. The loss function used in the calibration process is constructed by measuring the distance between historical and simulated stylised facts.

\subsubsection{Fat-tailed distribution of returns}
\hspace{0.3cm} The distributions of price returns have been found to be fat-tailed across all timescales. In other words, the return distributions exhibit positive excess kurtosis. Understanding positively kurtotic return distributions is important for risk management since large price movements are much more likely to occur than in commonly assumed normal distributions.

\hspace{0.3cm} Following the convention in literature, in this paper we investigate the stylised fact of fat-tailed returns by examining second-level intra-day price returns. Both millisecond-level historical and simulated mid-price time series are resampled into second-level frequency and we examine the mid-price returns for each second. Specifically, the last price snapshot is taken as the price for that specific second. Second-level price returns are calculated accordingly. Our experiments show that different time scales have no significant influence on the final results. The main metric used for evaluating the fat-tail characteristic is the Hill Estimator of the tail index (\citealt{hill2010tail}). A lower value of the Hill Estimator implies that the return distribution has a fatter tail.

\subsubsection{Autocorrelation of returns}
\begin{figure}[H]
\centering
\includegraphics[width=15cm, height=5cm, angle=0]{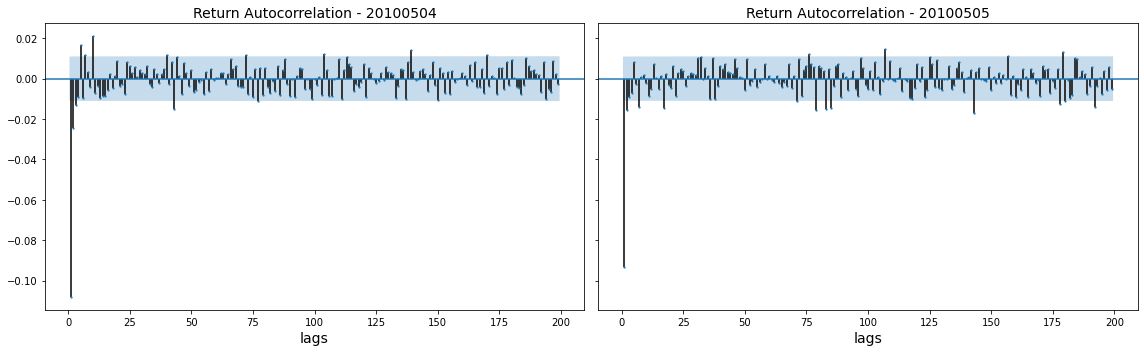}
\caption{Autocorrelation function of second-level returns for E-mini S\&P 500 futures on two days}
\label{acffigure}
\end{figure}

\hspace{0.3cm} Autocorrelation is defined to be a mathematical representation of the degree of similarity between a time series and a lagged version of the same time series. It measures the relationship between a variable's past values and its current value. Take first-order autocorrelation for example. A positive first-order autocorrelation of returns indicates that a positive (negative) return in one period is prone to be followed by a positive (negative) return in the subsequent period. Instead, if the first-order autocorrelation of returns is negative, a positive (negative) return will usually be followed by a negative (positive) return in the next period. It is observed that the returns series lack significant autocorrelation, except for weak, negative autocorrelation on very short timescales. \cite{mcgroarty2019high} show that the negative autocorrelation of returns is significantly stronger at a smaller time horizon and disappears at a longer time horizon. Examination of our data also reveals this stylised fact. Figure~\ref{acffigure} shows the autocorrelation function of second-level return time series for E-mini S\&P 500 futures on two days. We can see that the autocorrelation is significantly negative for very small lags, and the negative autocorrelation gradually disappears for larger lags.

\subsubsection{Volatility clustering}
\begin{figure}[H]
\centering
\includegraphics[width=16cm, height=5cm, angle=0]{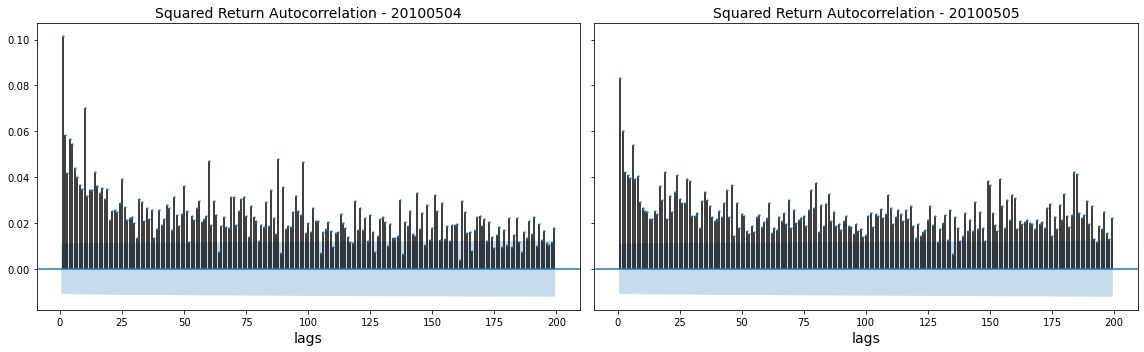}
\caption{Autocorrelation function of squared second-level returns for E-mini S\&P 500 futures on two days}
\label{squaredacffigure}
\end{figure}

\hspace{0.3cm} Financial price returns often exhibit the volatility clustering property: large changes in prices tend to be followed by large changes, while small changes in prices tend to be followed by small changes. This property results in the persistence of the amplitudes of price changes (\citealt{cont2007volatility}). It is found that the volatility clustering property exists on timescales varying from minutes to days and weeks. Volatility clustering also refers to the long memory of square price returns (\citealt{mcgroarty2019high}). Consequently, volatility clustering can be manifested by the slow decaying pattern in the autocorrelation of squared returns. Specifically, for short lags the autocorrelation function of squared returns is significantly positive, and the autocorrelation slowly decays with the lags increasing. Figure~\ref{squaredacffigure} shows the autocorrelation patterns for squared second-level returns for E-mini S\&P 500 futures on two days. It is shown that the volatility clustering stylised fact clearly exists in our collected E-mini S\&P 500 futures price dataset.

\subsubsection{Stylised Facts Distance as Loss Function} \label{lossfunction}
\hspace{0.3cm} The target for agent-based model calibration is to find an optimal set of model parameters to make the model generate a realistic simulated financial market. To solve this optimization problem, it is essential to have a metric that is able to quantify the "realism" of a simulated financial market. First of all, a realistic simulated financial market must exhibit similar characteristics to real financial markets, such as the fat-tailed return distribution and volatility level. In addition, realistic simulated financial data are also required to reproduce other stylised facts such as the autocorrelation patterns in returns and squared returns. Here we design a stylised facts distance to quantify the similarities between simulated and historical financial data. Four metrics are considered in the stylised facts distance: Hill Estimator of the tail index for absolute return distributions, volatility, autocorrelation of returns and autocorrelation of squared returns. For each metric, the differential quantity between simulated value and historical value is calculated. The stylised facts distance is then calculated as the weighted sum of the four differential quantities:

\begin{equation} \label{rawdistance}
\begin{aligned}
D &= w_1 * \Delta_{Hill} + w_2 * \Delta_{V} + w_3 * \Delta_{ACF^1} + w_4 * \Delta_{ACF^2} \\
\end{aligned}
\end{equation}

Detailed calculations of the four quantities in the stylised facts distance are presented below.

\hspace{0.3cm} The Hill Estimator is famous for inferring the power behaviour in tails of experimental distribution functions. Following \cite{FRANKE20121193}, we use the Hill Estimator of the tail index to estimate the degree of fat-tail in the distribution of absolute returns on the mid-price. Note that the absolute return distribution is considered since there is no need to distinguish between extreme positive and negative returns. In our experiments, the Hill Estimator for simulated absolute returns and historical absolute returns are calculated, respectively. The absolute difference between the two Hill Estimators constitutes the first part of the stylised facts distance:

\begin{equation} \label{hillestimator}
\begin{aligned}
\Delta_{Hill} &=  |Hill_s - Hill_h|\\
\end{aligned}
\end{equation}

where $Hill_s$ and $Hill_h$ are the Hill Estimator for simulated absolute return distribution and historical absolute return distribution, respectively. Intuitively, this part of the stylised facts distance quantifies the distance between simulated and historical return distribution, in terms of the heavy-tailedness. It addresses the requirement that the simulated return distribution exhibits the fat tail property, as is the case in historical return distribution.

\hspace{0.3cm} The second part of the stylised facts distance is the absolute volatility difference between simulated returns and historical returns:
\begin{equation} \label{voldiff}
\begin{aligned}
\Delta_{V} &=  |V_s - V_h| \\
\end{aligned}
\end{equation}
where $V_s$ and $V_h$ denote simulated volatility and historical volatility, respectively. This part addresses the requirement that a simulated financial market should be similar to real market in terms of volatility.

\hspace{0.3cm} The third part of the stylised facts distance is the difference between simulated and historical autocorrelations of returns. This part in the stylised distance measures the model's ability to reproduce autocorrelation patterns commonly found in historical returns. It is shown that financial price return time series lack significant autocorrelation, except for short time scales, where significantly negative autocorrelations exist. This phenomenon is backed by our empirical data. For very small lags the autocorrelations are negative, while for larger lags the autocorrelations become insignificant. To measure the distance in autocorrelation patterns between simulated data and historical data, we invoke the autocorrelation function of returns and calculate the average absolute difference between autocorrelations of simulated return time series and historical return time series for various lags:
\begin{equation} \label{acf1diff}
\begin{aligned}
\Delta_{ACF^1} &=  \frac{ \sum\limits_{l \: in \: lags} |ACF_s(l, r) - ACF_h(l, r)| } { |lags|} \\
\end{aligned}
\end{equation}

where $ACF_s(l, r)$, $ACF_h(l, r)$ are the autocorrelation function of lag $l$ for simulated returns and historical returns, respectively. $|lags|$ denotes the number of lags used in the calculation. Because the empirical autocorrelations are negative for very small lags and close to zero for larger lags, it is not necessary to consider all of the autocorrelation coefficients. Empirical evidence suggests that the autocorrelation pattern is well represented by the coefficients for three lags: 30, 60, and 90. Also, to reduce the effects of accidental outliers, the autocorrelation function is smoothed by calculating the three-lag average. That is, the lag 30 autocorrelation is calculated as the average autocorrelation of lag 30, 31, 32, and so is the calculation for lag 60 and lag 90. In total, autocorrelations of 9 lags (30, 31, 32, 60, 61, 62, 90, 91, 92) are considered and included in the calculation.

\hspace{0.3cm} The last part of the stylised facts distance is the difference between simulated and historical autocorrelations of squared returns. The replication of autocorrelation patterns in squared returns indicates the model's capability to reproduce the volatility clustering stylised fact. It is shown empirically that large price changes tend to be followed by other large price changes, known as the volatility clustering phenomenon. Consequently, though there are generally no significant patterns in autocorrelations of returns, the autocorrelations of squared returns are significantly positive, especially for small time lags. Also, as time lag increases, the autocorrelation of squared returns displays a slowly decaying pattern, as shown in Figure~\ref{squaredacffigure}. Similar to the difference between autocorrelations of returns $\Delta_{ACF^1}$, the difference between autocorrelations of squared returns is calculated as follows:

\begin{equation} \label{acf2diff}
\begin{aligned}
\Delta_{ACF^2} &=  \frac{ \sum\limits_{l \: in \: lags} |ACF_s(l, r^2) - ACF_h(l, r^2)| } { |lags|} \\
\end{aligned}
\end{equation}

where $ACF_s(l, r^2)$, $ACF_h(l, r^2)$ are the autocorrelation function of lag $l$ for simulated squared returns and historical squared returns, respectively. $|lags|$ denotes the number of lags used in the calculation. Here we use a slightly different $lags$ from the case for autocorrelations of returns calculation. Since empirical autocorrelations of squared returns are significantly positive and slowly decaying, we consider the autocorrelations of squared returns for four lags: 1, 30, 60, and 90. The three-lag average smoothing is also applied in the calculation. In total, the autocorrelations of 12 lags (1, 2, 3, 30, 31, 32, 60, 61, 62, 90, 91, 92) are considered and included in the calculation.

\hspace{0.3cm} The above four parts, along with the corresponding weights, constitute the stylised facts distance in Equation~(\ref{rawdistance}). The next question is how to determine the associated weight for each part of the stylised facts distance. The basic guiding idea is that the higher the sampling variability of a given part in historical data, the larger the difference between simulated value and historical value that can still be deemed insignificant. A natural candidate for each weight is the inverse of the sampling variance for the corresponding part in the stylised facts distance:

\begin{equation} \label{distanceweight}
\begin{aligned}
w_i &= \frac{1}{\sigma^2_i}  \\
\end{aligned}
\end{equation}

Following \cite{FRANKE20121193}, the sampling variance $\sigma^2_i$ for each part in the stylised facts distance is estimated by applying the block bootstrap method on the historical return time series.

\hspace{0.3cm} Note that the stylised facts distance is a function of model parameters. In other words, given a set of model parameters, there is a unique stylised facts distance calculated from the simulated time series, which corresponds to that particular set of model parameters. Let $\bm{\theta}$ denote the vector of model parameters to be estimated, Equation~(\ref{rawdistance}) can be rewritten as:
\begin{equation} \label{distancefunction}
\begin{aligned}
D(\bm{\theta}) &= w_1 * \Delta_{Hill}(\bm{\theta}) + w_2 * \Delta_{V}(\bm{\theta}) + w_3 * \Delta_{ACF^1}(\bm{\theta}) + w_4 * \Delta_{ACF^2}(\bm{\theta}) \\
\end{aligned}
\end{equation}
The smaller $D(\bm{\theta})$ is, the more realistic the simulation is. Thus $D(\bm{\theta})$ serves as the loss function that the calibration method aims to minimize by finding an optimal set of model parameters. Let $\bm{\Theta}$ denotes the admissible set for model parameter vector $\bm{\theta}$, the calibration target is to find the optimal model parameter vector $\bm{\hat{\theta}}$ that minimizes the stylised facts distance:
\begin{equation} \label{thetahat}
\begin{aligned}
\bm{\hat{\theta}} &= \arg \; \min_{\bm{\theta} \in \bm{\Theta}} \; D(\bm{\theta}) \\
\end{aligned}
\end{equation}

\subsection{Calibration Workflow and Results}
\hspace{0.3cm} The model is calibrated by choosing values for the model parameters so that the dynamics of simulated price time series match those observed empirical price time series. Specifically, the aim of the calibration process is to find $\bm{\hat{\theta}}$ that minimizes the stylised facts distance, as specified in Equation~\ref{thetahat}. As mentioned, data used in the calibration are the price time series data of the most liquid contract of E-mini S\&P 500 futures, from May 3rd, 2010 to May 6th, 2010. The model parameters are calibrated for every trading day\footnote{Due to the flash crash event in the afternoon trading session on May 6th, 2010, only the first half of the trading data (8:00-12:30) is used to calibrate the model parameters for this trading day.}. After a large-scale trial and error, seven model parameters are selected to be calibrated. The seven parameters are: $\mu_{\ell}$, $\sigma_{NT}$, $\kappa_1$, $\kappa_2$, $\beta_L$, $\beta_S$, $\theta_{MM}$. Specific meanings of these seven parameters are presented in Section~\ref{modelstructure}. In terms of the replication of stylised facts, our experiments show that the simulation results are less sensitive to the values for other parameters. As a result, it is reasonable to keep other parameters fixed. This choice significantly reduces the computational complexity of the calibration process. These parameters and corresponding descriptions, as well as specific values, are presented in Appendix~\ref{modelparameterdescription} and Appendix~\ref{fixedparameters}.

\hspace{0.3cm} The focus of this paper is not on specific methods for calibration; instead, we will pay more attention to the validation part. Here we briefly present the calibration process. The calibration workflow has two stages in our experiments. The main calibration technique in the first stage is the surrogate modelling approach, proposed by \cite{LAMPERTI2018366}. Specifically, an XGBoost surrogate model is built to approximate the agent-based model simulation. The surrogate model is capable of intelligently guiding the exploration of the parameter space. Estimated parameter values are those that give rise to smaller stylised facts distance, indicating that simulated moments match those observed empirically. After the first stage of the calibration, optimal parameters are given by the surrogate model. Even though global optimum is not guaranteed, it is shown in our experiments that the obtained parameter combination yields small stylised facts distance and is capable of reproducing realistic price dynamics. It is likely that the global optimal parameters are located close to the parameters generated by the first stage. Taking this into consideration, a numerical grid search over a feasible bounded set of parameters is carried out. The feasible set of parameters is centered around the optimal parameters given by the surrogate modelling approach in stage one. The parameter combination that yields the smallest stylised facts distance is selected as the final calibrated model parameter combination. Calibrated model parameter values for each trading day, as well as the yielded stylised facts distance, are presented in Table~\ref{calibratedvalues}.

\begin{table}[H]
\centering
\caption{Calibrated values for the seven model parameters and the yielded stylised facts distances}
\begin{tabular}{ | m{1.5cm}<\centering | m{1.2cm} | m{1.2cm} | m{1.2cm} | m{1.2cm} | m{1.2cm} | m{1.2cm} | m{1.2cm} | m{1.2cm} |} 
\hline
Date & $\mu_{\ell}$ & $\sigma_{NT}$ & $\kappa_1$ & $\kappa_2$ & $\beta_L$ & $\beta_S$ & $\theta_{MM}$ & $D(\bm{\hat{\theta}})$\\
\hline
20100503 & 0.9093  & 0.7895 & 0.1632 & 0.0235 & 0.0924 & 0.0252 & 0.7805 & 0.2259\\ 
\hline
20100504 & 1.2717  & 0.6732 & 0.0933 & 0.3402 & 0.6964 & 0.681 & 0.8998 & 0.2269\\ 
\hline
20100505 & 1.7396  & 0.6176 & 0.2844 & 0.2098 & 0.5935 & 0.055 & 0.8291 & 0.2191\\ 
\hline
20100506 & 1.0098  & 0.0107 & 0.3697 & 0.1649 & 0.7494 & 0.9877 & 0.3265 & 0.1666\\ 
\hline
\end{tabular}
\label{calibratedvalues}
\end{table}

\begin{figure}[H]
\centering
\includegraphics[width=16cm, height=5cm, angle=0]{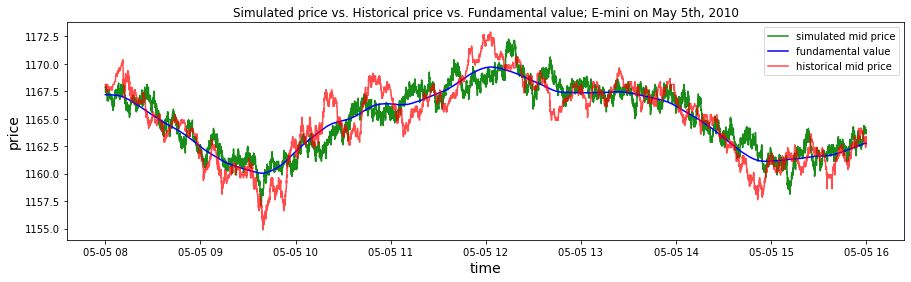}
\caption{Comparison between empirical mid-price time series on May 5th and simulated mid-price time series on May 5th}
\label{simulationexample}
\end{figure}

\hspace{0.3cm} Figure~\ref{simulationexample} compares empirical time series of mid-price on May 5th to simulated mid-price time series. Visual inspection shows that the model produces price time series whose dynamics are very similar to those in empirical data. Nevertheless, quantitative assessment is required to validate the proposed simulation model, which is presented in the subsequent section.

\subsection{Model Validation}
\hspace{0.3cm} Table~\ref{calibratedvalues} shows the stylised facts distance of the calibrated model. However, the value itself does not present an intuitive description of how well the simulated data fit empirical data. A cross-check on the validity of our model topology and calibration strategy is needed. Following \cite{FRANKE20121193}, two metrics are used for assessing the quality of the moment matching and the validity of the model simulation: the moment-specific p-value and the moment coverage ratio.

\subsubsection{Statistical Hypothesis Testing: Moment-specific p-value}
\hspace{0.3cm} The stylised facts distance provides us with numerical values for the realism of the simulation. However, one more elementary question needs to be addressed: whether the data generated by model calibration and simulation would be rejected by the empirical data. The question is answered by calculating a moment-specific p-value as a statistical hypothesis test. 

\hspace{0.3cm} Recall that in Section~\ref{lossfunction} the weights are calculated by the block bootstrapping method proposed in \cite{FRANKE20121193}. While the variance of block bootstrapping samples is the corresponding weight for each part of the loss function, the large number of samples obtained in this procedure can also be used to apply the loss function $D$ to them. In this way an entire frequency distribution of values of $D$ is available, which can subsequently be contrasted with the simulated distribution of values of $D$. The fundamental idea is that the set of bootstrapped samples of return time series is a proxy of the set of different return time series samples that could be produced if the hypothetical real-world data generation process exists. Accordingly, if a simulated return time series yields a value of $D$ within the range of the bootstrapped values of $D$, this simulated return time series is difficult to be distinguished from a real-world series. 

\hspace{0.3cm} Note that a Monte Carlo experiment is undertaken and the simulation is repeated many times to create a distribution of model-generated $D$ values. In total, two distributions of $D$ values are obtained: bootstrapped distribution and Monte Carlo simulated distribution. Let $N$ denote the number of bootstrapped samples. For consistency, we also select $N$ Monte Carlo samples to create the simulated distribution. Let $D^b$ and $D^m(\bm{\hat{\theta}})$ denote the bootstrapped sample and Monte Carlo sample, respectively\footnote{The Monte Carlo simulated distance value $D$ is dependent on the estimated set of parameters $\bm{\hat{\theta}}$.}. The distributions of the following two sets of $D$-values are then compared:

\begin{equation} \label{twodistribution}
\begin{aligned}
Bootstrap &: \{D^b\}_{b=1}^{N}  \\
Simulated &: \{D^m(\bm{\hat{\theta}})\}_{m=1}^{N}
\end{aligned}
\end{equation}

The p-value is obtained as follows. A critical value of $D$ is established from the bootstrapped distribution of values of $D$. Taking account of the rare extreme events with a significance level of 5\%, the critical value is defined to be the 95\% percentile $D_{0.95}$ of the bootstrapped $D$. In this way, a simulated return time series will be considered to be inconsistent with the real-world data, and therefore be rejected, if the corresponding $D$-value is larger than $D_{0.95}$. Take May 3rd trading day as an example. Here $D=0.2114$ is obtained as the critical value of the bootstrapped distance function. As for the Monte Carlo simulation, the model-generated distribution of $D$ is prominently wider to the right. Detailed calculation yields that $D_{0.95}$ corresponds to the 18.33\% percentile of the simulated distribution of values of $D$. Following the general moment matching convention in \cite{FRANKE20121193}, the model is said to have a p-value of 0.1833, with respect to the estimated parameter vector $\bm{\hat{\theta}}$, the historical data for May 3rd, and the specific moments that we have chosen. According to the conventional significance criteria, for this trading day the model will not be rejected as being obviously inconsistent with the empirical data. Following \cite{FRANKE20121193}, a p-value of 0.1833 is actually believed to be a fairly good performance for an agent-based financial market simulator. The p-values for the calibrated model for all trading days are shown in Table~\ref{pvaluetable}.

\begin{table}[H]
\centering
\caption{P-values for the calibrated model for all trading days}
\begin{tabular}{ | m{2.5cm}<\centering | m{2.5cm}<\centering | m{2.5cm}<\centering | m{2.5cm}<\centering | m{2.5cm}<\centering |} 
\hline
Date & 20100503 & 20100504 & 20100505 & 20100506\\
\hline
p-value & 0.1833  & 0.4167 & 0.0833 & 0.6167 \\ 
\hline
\end{tabular}
\label{pvaluetable}
\end{table}

\hspace{0.3cm} From Table~\ref{pvaluetable} we can see that for all trading day in the collected dataset, the moment-specific p-values are greater than 0.05. Consequently, we cannot reject the null hypothesis which specifies that the simulated return time series belong to the same distribution as the empirical return time series. This statistical hypothesis testing gives evidence that the calibrated model is capable of generating realistic financial price time series.

\subsubsection{Moment Coverage Ratio}
\hspace{0.3cm} The previous evaluation of the model was based on the values of the stylised facts distance function $D$. While statistical testing enables us to evaluate the validity of the model simulation, the quality of moment matching for each specific moment is still unknown. Another potential issue is that the stylised facts distance function $D$ is the optimisation target during the calibration process. Thus the evaluation metric involving the distance $D$ may be biased because of the potential overfitting problem. To address the above problems, the "moment coverage ratio" (MCR) metric is adopted to assess the degree of moment matching, taking into account each specific moment. The moment coverage ratio is originally proposed in \cite{FRANKE20121193}. The basis for moment coverage ratio calculation is the concept of a confidence interval of the empirical moments. Consistent with \cite{FRANKE20121193}, the 95\% confidence interval of a moment is considered, which is defined to be the interval with boundaries $\pm$1.96 times the standard deviation around the empirical value of this moment. The next step is to determine the standard deviation for each empirical moment. \cite{FRANKE20121193} apply the delta method to the autocorrelation coefficients to calculate the standard deviation. In this paper, we use a more direct way to obtain the empirical standard deviation, which is based on the block bootstrapping method. Recall that large quantities of return time series are obtained by the block bootstrapping method. For each specific moment, a moment value can be calculated out of every sampled return series. In total there will be $B$ values for each moment, where $B$ is the bootstrapping sample number. The standard deviation of those values is considered to be the standard deviation for the corresponding empirical moment. One may feel uneasy about the bootstrapping of the autocorrelation functions at the longer lags since the method alters the temporal order of the return series. However, the block size in our block bootstrapping method is 1800, which is significantly larger than the longest lag (90) in the autocorrelation functions. Consequently, the impact of bootstrapped block re-ordering on the autocorrelation functions is negligible. 

\hspace{0.3cm} With the standard deviation on hand, the corresponding confidence interval for each specific moment is immediately available. In this way an intuitive criterion for assessing a simulated return series is obtained: if all of its moments are contained in the confidence intervals, the simulated return series cannot be rejected as being incompatible with empirical data. Nonetheless, one single simulation is not sufficient to evaluate a model as a whole due to the sample variability. In addition, it is likely that for one simulated return series some moments are contained in the confidence interval while others are not. It goes without saying that considering multiple simulation runs of the model will provide a more exhaustive assessment of the model performance. Specifically, for each simulation run the confidence interval check is repeated. We count the number of Monte Carlo simulation runs in which the single moments are contained in the corresponding confidence intervals. The corresponding percentage numbers out of all Monte Carlo runs are defined as the moment coverage ratio. 

\hspace{0.3cm} Since the model is calibrated by each trading day, Monte Carlo simulation is run for each trading day and the corresponding moment coverage ratios are calculated to evaluate the calibrated model. Table~\ref{mcrtable} presents the results of the moment coverage ratios calculation. Except for the volatility moment on May 3rd and the 90-lag squared return autocorrelation moment on May 4th, all other moment coverage ratios are higher than 50\%. According to the analysis in \cite{FRANKE20121193}, the higher than 50\% moment coverage ratio represents a terrific performance of the model. In addition, almost half of the moment coverage ratios are even higher than 90\%, which indicates that our calibrated model has an excellent ability to reproduce realistic stylised facts. Overall, with respect to the selected moments, the calibrated model's capability of matching empirical moments and reproducing realistic stylised facts is highly remarkable.

\begin{table}[H]
\centering
\caption{Moment coverage ratios of the calibrated model}
\begin{tabular}{ | m{3.5cm}<\centering | m{2.5cm}<\centering | m{2.5cm}<\centering | m{2.5cm}<\centering | m{2.5cm}<\centering |} 
\hline
\backslashbox{Moment}{Date} & 20100503 & 20100504 & 20100505 & 20100506\\
\hline
Hill estimator (inverse) & 98.3\% & 66.7\% & 98.3\% & 100.0\% \\ 
\hline
Volatility & 0.0\% & 85.0\% & 100.0\% & 56.7\% \\ 
\hline
Return autocorrelation (30 lags) & 85.0\% & 78.3\% & 50.0\% & 90.0\% \\ 
\hline
Return autocorrelation (60 lags) & 100.0\% & 98.3\% & 96.7\% & 96.7\% \\ 
\hline
Return autocorrelation (90 lags) & 98.3\% & 60.0\% & 100.0\% & 58.3\% \\ 
\hline
Squared return autocorrelation (1 lag) & 100.0\% & 98.3\% & 95.0\% & 93.3\% \\ 
\hline
Squared return autocorrelation (30 lags) & 78.3\% & 51.7\% & 71.7\% & 98.3\% \\ 
\hline
Squared return autocorrelation (60 lags) & 88.3\% & 10.0\% & 80.0\% & 73.3\% \\ 
\hline
Squared return autocorrelation (90 lags) & 91.7\% & 46.7\% & 76.7\% & 51.7\% \\ 
\hline
\end{tabular}
\label{mcrtable}
\end{table}


\section{2010 Flash Crash Scenarios} \label{flashcrashscenarios}

\hspace{0.3cm} As a real-world application, the proposed model is used to investigate market dynamics during flash crash events and the conditions for the occurrence of flash crash scenarios. This section will present the reproduction and investigation of a famous historical flash crash event - the flash crash on May 6th, 2010. In the next section, we will investigate conditions for the occurrence of mini flash crash events.

\subsection{Simulating Historical Flash Crash}
\hspace{0.3cm} We simulate the 2010 Flash Crash within the framework of our high-frequency financial market simulator. The flash crash happened in the afternoon trading session on May 6th, 2010, starting at around 14:30. As mentioned before, the model parameters are calibrated to the data in the morning trading session (8:00-12:30) to avoid overfitting\footnote{Except for market maker inventory limit and fundamental trader trading frequency, which are key to generating realistic flash crashes.}. According to the CFTC-SEC staff report, an automated execution algorithm, which aimed at selling a large number of contracts, was identified as one important trigger for the flash crash. Consistent with this report, an institutional trader is introduced in the simulator to mimic the automatic Sell Algorithm. The institutional trader will initiate a Sell Algorithm that intends to sell a large quantity of contracts at 14:30 in our simulator. The parameters associated with the institutional traders are tuned to reproduce realistic flash crash behaviours as in historical data. We present the trading configuration of the institutional trader in the subsequent session, followed by a detailed analysis of the market dynamics during the simulated flash crash event.

\subsubsection{Introduction of Institutional Trader (INS)}
\hspace{0.3cm} The institutional trader is specially designed to replicate the behaviour of the large institutional trader who initiated the sell program to sell a large quantity of E-mini contracts on May 6th, 2010. In the beginning, the institutional trader has a large inventory of $Q$ E-mini contracts. The institutional trader will start a sell program to sell the inventory at 14:30 in the simulation. Consistent with the historical scenarios, the sell program in the simulation is executed via an automated execution algorithm that is programmed to submit orders into the market to target a certain execution rate, which is a percentage of the market trading volume calculated over the previous minute. The execution rate is denoted as $r$, which is set to 9\% in our simulation according to \cite{securities2010findings}. Note that the Sell Algorithm only considers total trading volume in the market, taking no account of the market price or time. This procyclical behaviour is potentially one causal factor in the occurrence of flash crash events. As for specific trading behaviours, the institutional trader always keeps count of the total trading volume $V$ of the last minute, and submits a market sell order every $n$ seconds. For each market order, the volume would be $V * r * \frac{n}{60}$, where $\frac{n}{60}$ is a scaling factor ensuring the target execution rate $r$ is achieved by the sell program. The institutional trader continues trading until all the inventory is cleared. Algorithm~\ref{inslogic} describes the trading behaviours of the institutional trader in detail\footnote{The $step\_seconds$ in the institutional trader logic is the number of seconds since the beginning of the simulation.}.

\begin{algorithm}
  \caption{Institutional Trader Logic}
  \label{inslogic}
  \begin{algorithmic}[1]
  \State Initialise $n$, $r$, $Q$
  \For{ \textbf{each} simulation step}
  \State $V \gets $ total trading volume in the market for the last minute
  \If{(simulation timestamp $ > 14:30$ ) $ \land (Q > 0) $ }
      \State $vol \gets V * r * \frac{n}{60}$;
      \If {($step\_seconds \mod n$) $ == 0$}
      \State Submit a market sell order with volume $vol$
      \State $Q \gets Q - vol$
      \Else {\\ \hspace{1.5cm} No action taken}
      \EndIf
  \Else{\\ \hspace{1.0cm} No action taken}
  \EndIf
  \EndFor
  \end{algorithmic}
\end{algorithm}

\subsubsection{Market Behaviours during Flash Crash}
\begin{figure}[H]
\centering
\includegraphics[width=16cm, height=5cm, angle=0]{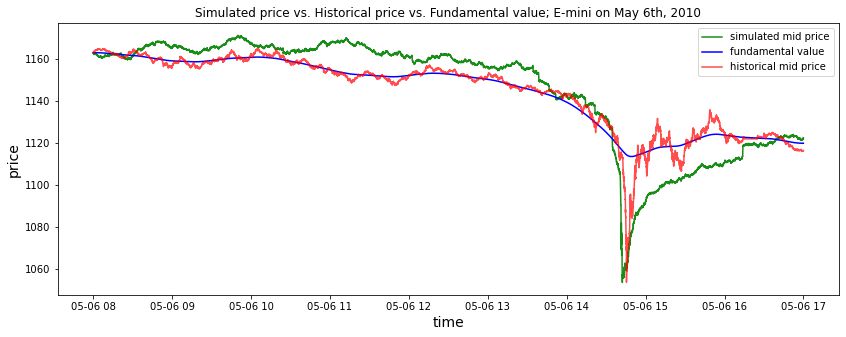}
\caption{One single simulation of the 2010 Flash Crash event. Red line is the historical price; green line is the simulated price; blue line is the fundamental value extracted by Kalman Smoother}
\label{simcrash}
\end{figure}

\hspace{0.3cm} To generate a realistic flash crash event, we fine-tune the parameters of the newly introduced institutional trader. According to \cite{securities2010findings}, the flash crash event is potentially caused by the "hot-potato" effect among high-frequency market makers and the mismatch of the trading frequency between different types of traders. Consequently, we also tune the market maker inventory limit and the trading frequency of fundamental traders. The effect of these parameters on the flash crash event will be presented in subsequent sessions. All other model parameters are fixed and are given exactly the same values that are calibrated to the morning trading session data on May 6th, 2010. This configuration helps to reduce the degree of freedom and avoid the potential problem of overfitting. The resulting simulation of the price time series mimics the real-world flash crash event. Figure~\ref{simcrash} presents such a single simulation of the price trajectory that undergoes a flash crash scenario. The corresponding parameter configuration is shown in Appendix~\ref{may6thparameterapp}. Figure~\ref{volume} presents a comparison between simulated trading volume and historical trading volume during the whole day. Carefully examining such a single simulation provides useful insights into how the trading behaviour of different types of traders interact to bring about a flash crash scenario. 

\begin{figure}[H]
\centering
\includegraphics[width=16cm, height=5cm, angle=0]{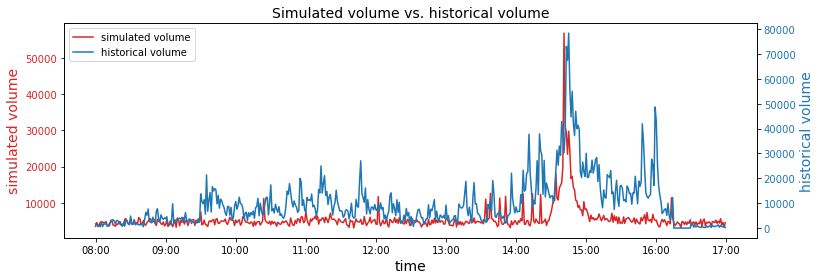}
\caption{Comparison between simulated volume and historical volume on May 6th, 2010}
\label{volume}
\end{figure}

\begin{figure}[H]
\centering
\begin{minipage}[t]{0.49\textwidth}
\centering
\includegraphics[width=8.1cm, height=4.5cm]{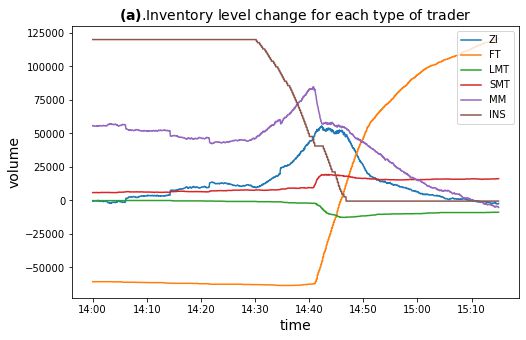}
\end{minipage}
\begin{minipage}[t]{0.49\textwidth}
\centering
\includegraphics[width=8cm, height=4.5cm]{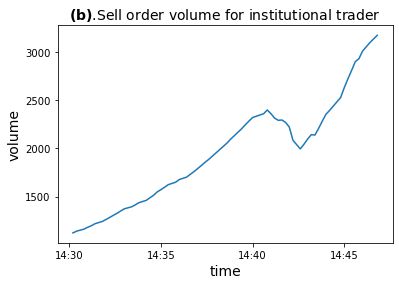}
\end{minipage}
\caption{\textbf{(a)}: Total inventory level for each type of trader, around the simulated flash crash event (14:00 - 15:15). \textbf{(b)}: The volume change for the selling orders of the institutional trader, during the time when institutional trader is active}
\label{inventories}
\end{figure}

\begin{figure}[H]
\centering
\includegraphics[width=16cm, height=8cm, angle=0]{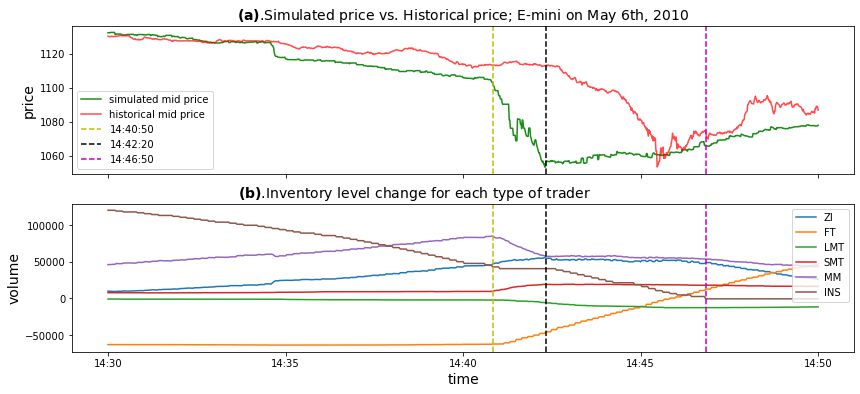}
\caption{Detailed simulated mid-price against the inventories of each category of trader from 14:30 to 14:50. \textbf{(a)}: Simulated and historical price from 14:30 to 14:50. \textbf{(b)}: Total inventory level for each type of trader during time period from 14:30 to 14:50}
\label{detailedpriceinventory}
\end{figure}

\begin{figure}[H]
\centering
\includegraphics[width=13cm, height=9.5cm, angle=0]{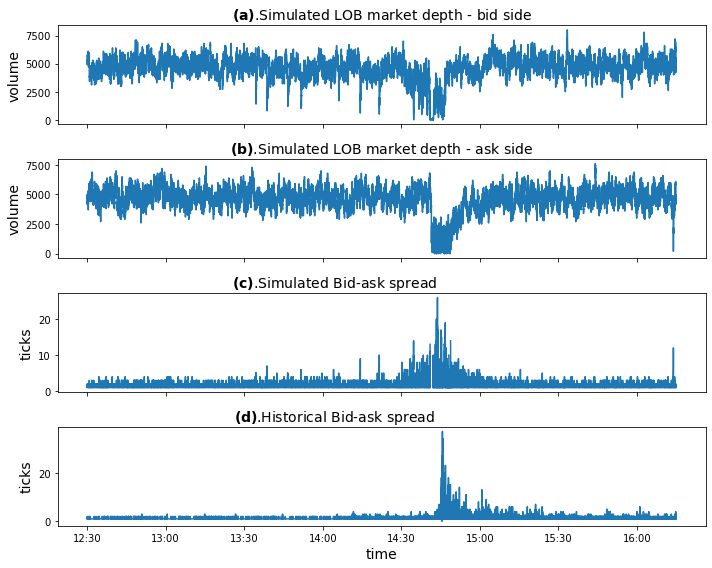}
\caption[]{\textbf{(a)}: Simulated market depth (total resting volume) of bid-side limit order book. \textbf{(b)}: Simulated market depth (total resting volume) of ask-side limit order book. \textbf{(c)}: Bid-ask spread in the simulated limit order book. \textbf{(d)}: Bid-ask spread in the historical limit order book. All four panels display time period spanning from 12:30 to 16:15}
\label{depthspread}
\end{figure}

\hspace{0.3cm} By visual inspection, the simulation accurately replicated the defining characteristics of a flash crash. Figure~\ref{inventories} presents the total inventory level for each type of trader around the simulated flash crash event, and the market sell order volume of the institutional trader. To assist with detailed inventory analysis during the flash crash event, Figure~\ref{detailedpriceinventory} presents the inventory level for each type of traders against simulated price during the time interval from 14:30 to 14:50. Figure~\ref{depthspread} displays the market depth for both sides of the simulated limit order book, and the comparison between simulated bid-ask spread and historical bid-ask spread. Detailed examination shows that the simulated flash crash accurately matches the market dynamics during the historical 2010 Flash Crash. The dynamics around the simulated flash crash event are as follows.

\begin{itemize}
\item The overall market sentiment in the simulation is negative, as we can see that the fundamental value is broadly decreasing before 14:00. At 14:00, according to Figure~\ref{inventories} the fundamental traders accumulate a negative inventory due to continuously decreasing fundamental value, while market makers accumulate a positive inventory. Before the flash crash event, the market depth for both sides of the limit order book is relatively stable, while the simulated bid-ask spread is at a relatively low level.
\item At 14:30 in the simulated market, the market price has already dropped by 2.63\% compared to the opening price level. Regardless of the downward market trend, the institutional trader initiates a sell program to sell a large number of inventories, which is 120,000 in our simulation. The institutional trader decides to execute the sell program via an automated execution algorithm, which is set to feed market selling orders into the market to target an execution rate of 9\%. That is, for each minute the institutional traders aim to sell 9\% of the market trading volume calculated over the previous minute. Neither price nor time is considered by the Sell Algorithm. Refer to Algorithm~\ref{inslogic} for specific trading logic of the institutional trader.
\item In our simulation, market makers correspond to the high-frequency traders in the real market. The above selling pressure is initially absorbed by market makers in our simulated market, which is consistent with the dynamics in the historical market (\citealt{securities2010findings}). This is shown by the inventory change for each type of traders in Figure~\ref{detailedpriceinventory}. From 14:30:00 to 14:40:50, the total inventory for market makers is rapidly increasing, while the total inventory for low-frequency fundamental traders barely changed. During this approximately 11-minute time period, the price undergoes a further 2.61\% drop. At 14:40:50, the market depth in the limit order book has slightly decreased and the bid-ask spread has slightly increased. The scale of change for the market depth and bid-ask spread is restricted because of the existence of market makers. 
\item In addition to the above observations, the market transacted volume is continuously increasing during the above 11-minute time interval due to the trading activities of the Sell Algorithm, as shown in Figure~\ref{volume}. Because of its procyclical nature, the Sell Algorithm used by the institutional trader responds to the increased volume by increasing the volume of sell orders that it is feeding into the market, even though the orders that it already submitted to the market were not yet fully absorbed and have caused non-trivial market turbulence. The increasing order volume of the institutional trader is shown in panel (b) of Figure~\ref{inventories}.
\item As market makers are the buyers of the initial batch of orders sent by the institutional trader, they have accumulated massive temporary long positions of the contracts. Nevertheless, the inventory of market makers cannot accumulate infinitely. At 14:40:50, the inventory limits for many market makers are exceeded. At this point, a dramatic and significant market crash starts. Those market makers, who have accumulated excessive positions than their inventory limits, stop providing liquidity and begin to aggressively sell their inventories in order to reduce their temporary long positions. This is consistent with the typical trading practice that a market maker tends to maintain a relatively small aggregate inventory for the purpose of risk management. Consequently, these market makers contribute further selling pressure to the market in addition to the Sell Algorithm. Still lacking sufficient demand from fundamental traders, the aggregate sell volume is consumed by remaining market makers who are still quoting in the market. As a result, more market makers accumulate excessive positions than their inventory limits, which in turn forces them to sell their long positions to the market. The same positions rapidly pass among all market makers. Such so-called "hot-potato" effect quickly sweep almost all market makers in the market, resulting in a dramatic price drop and quoting suspension of almost all market makers. 


\item The combined selling pressure from the Sell Algorithm and market makers drove the price down by more than 4.16\% in less than two minutes from 14:40:50 to 14:42:20, with the price reaching its intra-day low of 1053.5. Along with the price plunge, the market suffers from great liquidity loss.  The liquidity loss can be reflected by the sharp decrease of market depth during the simulated flash crash event. In the simulated morning trading session, the average market depth for both bid and ask sides are around 5000. In contrast, during the simulated large price plunge that starts at around 14:40, the market depth for both bid and ask sides are less than 1000, even reaches 0 in bid side for more than 1 minute. The significant liquidity loss in our simulation is mostly because of the withdrawal of market makers from the market. The withdrawal of market makers in our simulation is consistent with empirical findings in \cite{securities2010findings}, which states that real market makers did stop trading during the 2010 Flash Crash event. According to Figure~\ref{depthspread}, the market depth for both sides of the limit order book dropped dramatically during this time period. There is less than 1\% of bid-side market depth observed during normal trading hours, with even zero bid-side market depth for around 30 seconds in the middle of 14:41. The bid-ask spread widens dramatically, reaching more than 20 ticks at this short time interval. 

\item As price drops quickly, the demand from fundamental traders gradually increases according to Equation~\ref{fundamentalequation}. Figure~\ref{detailedpriceinventory} shows that fundamental traders do absorb a portion of selling volume during the sharpest price drop. However, the sudden decline in both price and liquidity indicates that the price was moving so fast that fundamental traders were incapable of providing enough buying support.

\item After the price hits the intra-day low level, the demand from fundamental traders finally increased to a level that is able to counteract the selling pressure from the institutional trader. Starting at 14:42:20, the price hovers at the lower level for approximately one minute and then starts to bounce back quickly towards the fundamental value. The Sell Algorithm continues to feed sell orders to the market until about 14:47, at which point the inventory of the institutional trader has been emptied.
\end{itemize}

\hspace{0.3cm} The above is a detailed analysis of the market dynamics of a simulated flash crash event. The simulated dynamics accurately match the dynamics of the historical 2010 Flash Crash event. Both simulated flash crash and historical flash crash have an amplitude of around 7\% (starting from 14:30:00), and both prices undergo a similar "flash crash" shape. A large Sell Algorithm is replicated to trigger the simulated flash crash, and is executed for around 17 minutes, which is close to the 20 minutes execution time of the historical Sell Algorithm. The overall liquidity loss during the flash crash is replicated accurately, with decreased market depth and enlarged bid-ask spread as emergent properties of the simulation. We also reproduce realistic patterns in the simulated market trading volume. The detailed progress of the simulated flash crash also matches the empirical analysis of the historical flash crash event.

\hspace{0.3cm} In the subsequent section we will explore the conditions that influence the severity of the flash crash event.

\subsection{Flash Crash Under Different Conditions} \label{conditionforlargeflashcrash}
\hspace{0.3cm} The above presents an analysis of the market dynamics during the simulated flash crash event. A natural question to follow is what factors contribute to the sharp price drop. One significant advantage of the agent-based model is the ability to be simulated multiple times under different conditions to reproduce realistic market events, while the historical event only happens once. This makes agent-based model a perfect testbed for exploring the conditions that lead to various rare market events, such as the flash crash. In this section, we explore the conditions that would influence the severity of flash crash events.

\hspace{0.3cm} To measure the severity of flash crash events, we define the amplitude of the flash crash event in our simulation as the maximum percentage price drop after 14:00. Specifically, the flash crash amplitude $Amp$ is calculated as below:
\begin{equation} \label{crashamplitude}
\begin{aligned}
Amp &= |\frac{TWAP_{14:00-14:05} - P_{min}}{TWAP_{14:00-14:05}}|
\end{aligned}
\end{equation}
where $TWAP_{14:00-14:05}$ is the time-weighted average price of time interval between 14:00 and 14:05, while $P_{min}$ is the lowest price during the flash crash simulation. 

\hspace{0.3cm} The main methodology applied here is Monte Carlo simulation experiments under controlled conditions. The conditions are represented by model parameters. Each combination of model parameters corresponds to a specific market condition configuration. For each combination of model parameters, multiple simulations are carried out. For each simulation, the amplitude of the simulated flash crash is calculated. In this way a distribution of flash crash amplitudes is obtained. To reduce the influence of extreme values, the 50\% quantile of all the Monte Carlo simulated amplitudes is defined as the flash crash amplitude associated with the condition that is represented by the specific combination of model parameters. The 40\% and 60\% quantile of the simulated flash crash amplitudes are also recorded. 

\begin{figure}[H]
\centering
\includegraphics[width=16.5cm, height=4cm, angle=0]{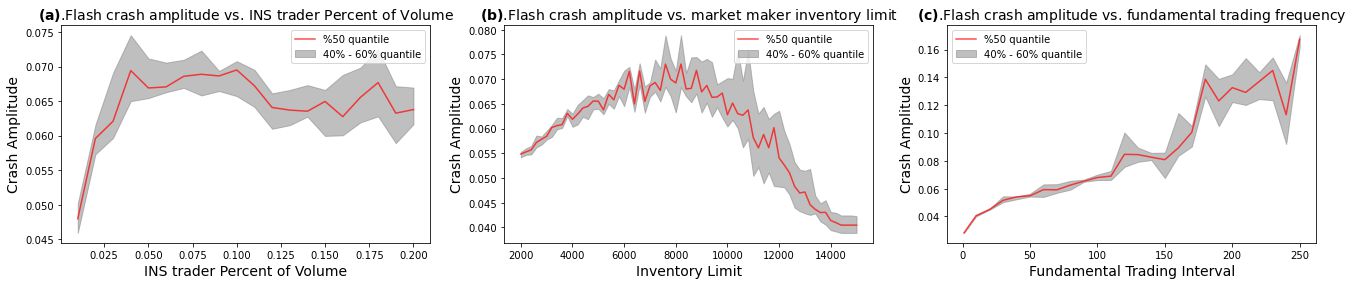}
\caption{\textbf{(a)}: Relationship between the amplitude of simulated flash crashes and the inventory limit of market makers. \textbf{(b)}: Relationship between the amplitude of simulated flash crashes and the percentage of volume $r$ for the Sell Algorithm (Institutional Trader). \textbf{(c)}: Relationship between the amplitude of simulated flash crashes and the trading frequency of fundamental traders. }
\label{amplimit}
\end{figure}

\hspace{0.3cm} To explore how one specific condition affects the severity of flash crash events, we change the corresponding model parameter and repeat the above Monte Carlo simulation to get the associated flash crash amplitude. The other model parameters are given the previous calibrated values and are kept fixed during this experimental process. This allows us to observe how the flash crash amplitude changes when only one specific model parameter changes. In particular, we are interested in three model parameters that relate closely to the flash crash amplitude: the percentage of volume for the Sell Algorithm ($r$), the inventory limit for each market maker ($\varepsilon^{MM}_{limit}$), and the trading frequency for fundamental traders ($FT_{freq}$). Figure~\ref{amplimit} presents the relationships between these model parameters and the simulated flash crash amplitude. Detailed analysis is as follows.

\paragraph*{Percentage of Volume for the Sell Algorithm \\}
\hspace*{0.3cm} One important parameter that characterises the Sell Algorithm is the percentage of volume $r$, which determines the target execution rate of the Sell Algorithm. Specifically, $r$ is the percentage of the market trading volume calculated over the previous minute that the Sell Algorithm aims to execute. Monte Carlo simulation is carried out with different values for $r$, while other model parameters are given the previous calibrated values and are strictly kept constant. Panel (a) in Figure~\ref{amplimit} shows the relationship between the simulated flash crash amplitude and the percentage of volume $r$.

\hspace{0.3cm} It is shown that when $r$ is small, the amplitude of the flash crash is limited. When $r$ is increased, the flash crash amplitude is also enlarged. Within a certain range, the more aggressive the Sell Algorithm (larger $r$), the more severe the flash crash event (larger amplitude). This phenomenon is consistent with our intuition. One interesting phenomenon is that after $r$ exceeds 5\%, the flash crash amplitude barely changes even though we continue to increase the value of $r$. The reason is that as long as the price reaches a certain lower level, the demand generated from fundamental traders is large enough to digest the inventory of the institutional trader. Since the total inventory of the institutional trader is fixed, the lowest price that could be reached during the simulation barely changed. Consequently, the flash crash amplitude stops increasing despite larger values for $r$.

\hspace{0.3cm} The lessons learned from this experiment mainly involves the choice of the algorithmic trading strategies for large institutional traders. The market impact of an algorithmic trading strategy may contradict our intuition, as is the case here when $r$ is larger than 5\%. Instead, the complex interaction between an algorithmic trading strategy and other market participants is likely to cause unexpected consequences. It is strongly recommended that institutional traders deploy agent-based financial market simulation to explore potential consequences before choosing a specific algorithmic trading strategy.

\paragraph*{Inventory limit for market makers \\}
\hspace*{0.3cm} Both empirical and simulation analyses indicate that high-frequency market makers play an important role in the flash crash event. Though market makers initially absorb lots of selling pressure from the institutional trader, they quickly turn into aggressive sellers after their inventory limits are exceeded, creating the "hot-potato" effect that exacerbates the crash. The inventory limit $\varepsilon^{MM}_{limit}$ plays an important role in controlling the significance of the "hot-potato" effect. Consequently, the value of $\varepsilon^{MM}_{limit}$ also influences the flash crash amplitude. We run Monte Carlo simulations with different values for $\varepsilon^{MM}_{limit}$. Other parameters are given the previous calibrated values and are kept constant. Panel (b) in Figure~\ref{amplimit} presents the relationship between the simulated flash crash amplitude and the inventory limit for each market maker $\varepsilon^{MM}_{limit}$.

\hspace{0.3cm} It is shown that the functional relationship between flash crash amplitude and market maker inventory limit is not monotonous. When the $\varepsilon^{MM}_{limit}$ is relatively small (smaller than 8000 in our simulation), the flash crash amplitude is an increasing function with regard to $\varepsilon^{MM}_{limit}$. This is because the larger the $\varepsilon^{MM}_{limit}$, the more inventory that the market makers will accumulate. With more accumulated inventory, the market makers will create much more selling pressure at the point when their inventory limits are exceeded, resulting in a more severe crash. However, the flash crash amplitude is decreasing on increasing the value of $\varepsilon^{MM}_{limit}$ if the $\varepsilon^{MM}_{limit}$ is larger than a certain level (around 8000 in our simulation). This is because the large enough $\varepsilon^{MM}_{limit}$ makes it possible for market makers to absorb a larger portion of selling pressure from the Sell Algorithm. As a result, there are fewer market makers whose inventory limits are reached, curtailing the "hot-potato" effect during the flash crash. In extreme cases when inventory limits for market makers are infinite, the market makers will absorb all selling pressure in the market and there will be no flash crash event.

\hspace{0.3cm} The analysis here emphasizes the importance of risk management for high-frequency market makers. Since all market makers have some sort of inventory control, the decreasing part of panel (b) in Figure~\ref{amplimit} is hardly feasible in the real-world trading environment. Instead, most real-world market makers have inventory limits that lie on the increasing part of the curve. An individual market maker may consider his own inventory limit to be proper; however, multiple "proper" inventory limits combined may lead to a significant crash under stressed scenarios due to the "hot-potato" effect. In order to foster a smooth and stable market, experimental results suggest that policymakers could impose certain inventory limits on all high-frequency market makers. 

\paragraph*{Trading frequency for fundamental traders \\}

\hspace*{0.3cm} According to \cite{karvik2018deeds}, the mismatch of trading frequency between different types of traders is an important factor that leads to flash crash scenarios. The flash crash happens when the market is dominated by the procyclical behaviours of high-frequency market participants. The price moves so fast that lower frequency fundamental traders are unable to supply enough buy-side liquidity. So what will happen if we change the relative frequency between the market makers and fundamental traders in our simulation? Experiments are carried out with different trading frequencies for fundamental traders. The trading frequency of fundamental traders is changed by changing $S^{FT}_{interval}$, which is the interval between actions from the same fundamental traders. Specifically, once a fundamental trader submits an order to the market, the same fundamental trader is not able to participate in the simulated market within the subsequent $S^{FT}_{interval}$ steps. Only after $S^{FT}_{interval}$ step can this fundamental trader submit another order to the market. In the default calibrated simulation $S^{FT}_{interval}$ is 100, while the same interval for market makers is 1. In other words, market makers can act in every simulation step, while a fundamental trader can only act once in every 100 steps. The smaller the $S^{FT}_{interval}$ is, the higher the trading frequency for fundamental traders. If $S^{FT}_{interval}$ has value 1, the fundamental traders would have the same trading frequency as market makers, creating the imaginary "high-frequency fundamental traders". Panel (c) in Figure~\ref{amplimit} presents how the flash crash amplitude changes when the trading frequency of fundamental traders changes.

\hspace{0.3cm} The figure indicates that there is a monotonous relationship between flash crash amplitude and the trading frequency of fundamental traders. Note that the larger the interval is, the lower the trading frequency is. Thus the flash crash amplitude is decreasing when we increase the trading frequency of fundamental traders. If the fundamental traders have the same trading frequency as the market makers, the flash crash amplitude becomes small enough that the flash crash turns into a small shock in the market. 

\hspace{0.3cm} The analysis here supports the argument that the mismatch of trading frequency between different types of traders potentially leads to flash crash events. There is hardly an imaginary "high-frequency fundamental trader" in the real-world trading environment. Thus practical trading environment corresponds to the right-hand part of the functional relationship in Panel (c) of Figure~\ref{amplimit}, where fundamental traders act at a lower frequency. The results indirectly show that high-frequency trader is an important factor in the occurrence of flash crash scenarios.

\section{Mini Flash Crash Scenarios} \label{miniflashcrashscenarios}
\hspace{0.3cm} The above section provides an analysis of the occurrence of flash crash events. The 2010 Flash Crash is so large that no following events have rivalled its depth, breadth, and speed of price movement (\citealt{paddrik2017effects}). However, flash crashes on a smaller scale do occur more frequently. \cite{johnson2012financial} identify more than 18,000 mini flash crash incidents between 2006 and 2011 in US equity markets. Those mini flash crashes are characterised to be abrupt and severe price changes over a short period of time. One natural question is how those mini flash crashes happened. As a second application for the proposed agent-based model, we investigate and analyse the causes of mini flash crashes in the framework of our agent-based financial market simulation. An innovative "Spiking Trader" is introduced to the market to mimic the trigger of mini flash crash events. A typical mini flash crash event in our simulation is analysed in detail, followed by experiments for exploring the conditions for mini flash crash events.

\subsection{Introduction of Spiking Trader (ST)}
\hspace{0.3cm} In the current literature, there is a growing consensus that the mini flash crashes result from interactions between various trading algorithms that operate at or beyond the limits of human response times, such as the procyclical behaviours of high-frequency market participants. However, there are various triggers for mini flash crash events. According to \cite{karvik2018deeds}, one specific trigger could be orders that are large relative to the supply of available limit orders, which bring price shocks to the market. To mimic the occasional price shock in the market, a type of trader agent called "Spiking Trader" is introduced to our agent-based artificial financial market.

\hspace{0.3cm} The proposed spiking trader mainly creates price shocks to the market, either upward or downward. Initially the spiking trader is inactive. For each simulation step, there is a small probability that a spiking trader is activated. Once activated, the spiking trader will submit market orders of the same direction for several consecutive simulation steps, creating price shocks in one direction. A spiking trader is associated with three parameters: $N_{spike}$, $\mu_{spike}$, and $V_{spike}$. $N_{spike}$ represents the number of consecutive orders to be submitted after the spiking trader is activated. $\mu_{spike}$ is the probability of being activated in a certain simulation step. $V_{spike}$ is the volume of the orders from spiking traders. Detailed trading logic for spiking traders is shown in Algorithm~\ref{spikelogic}. Note that $N_{ST}$ spiking traders are introduced to the agent-based simulation model. In our experiments, $N_{ST}$ has value 2.

\begin{algorithm}
  \caption{Spiking Trader Logic}
  \label{spikelogic}
  \begin{algorithmic}[1]
  \State Initialise $N_{spike}$, $\mu_{spike}$, $V_{spike}$
  \State $Status \gets 0$; $D \gets $ Sell
  \For{ \textbf{each} simulation step}
  
  \If{$Status > 0$ }
    \State Submit a market order with volume $V_{spike}$, direction $D$
    \State $Status \gets Status - 1$
  \Else
      \If {$p \in U(0, 1) < \mu_{spike}$}
        \State \textbf{if} $p \in U(0, 1) < 0.5$ \textbf{then} $D \gets $ Sell  \textbf{else} $D \gets $ Buy \textbf{end if}
        \State $Status \gets N_{spike}$
      \Else {\\ \hspace{1.5cm} No action taken}
      \EndIf
  \EndIf
  \EndFor
  \end{algorithmic}
\end{algorithm}

\subsection{Mini Flash Crash Analysis} 
\hspace{0.3cm} To investigate mini flash crash events, there are some modifications to the simulation configuration. The number of market makers in the simulation is reduced. Our experiments show that this increases the probability of the occurrence of mini flash crash events. The reason is that less market makers will lead to thinner market depth, generating more mini flash crash events for scrutiny. The institutional trader is removed from the market; while two spiking traders are introduced to the market simulation. Other model parameters are the same as the simulation for the 2010 Flash Crash. Detailed model parameters for mini flash crash simulation are shown in Appendix~\ref{miniparameterapp}.

\begin{figure}[H]
\centering
\includegraphics[width=15cm, height=5cm, angle=0]{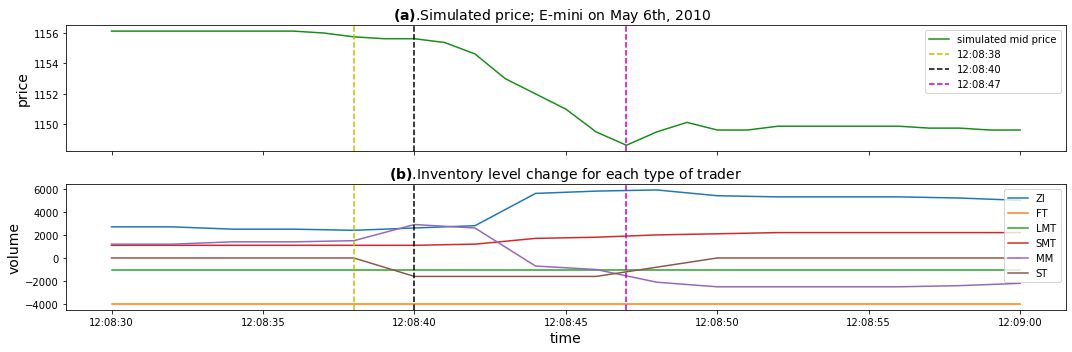}
\caption{Detailed simulated price against the inventories of each type of trader during the simulated mini flash crash event. \textbf{(a)}: Simulated and historical price from 12:08:30 to 12:09:00. \textbf{(b)}: Total inventory level for each type of trader during time period from 12:08:30 to 12:09:00}
\label{minidetailed}
\end{figure}

\hspace{0.3cm} Figure~\ref{miniflashcrash} presents a typical mini flash crash scenario in our simulation. The price drops nearly 80 basis points in just several seconds, and then bounces back towards the fundamental value. Figure~\ref{minidetailed} shows the inventory level for each type of trader against simulated price during the mini flash crash scenario. According to Figure~\ref{minidetailed}, the trigger and dynamics for that specific simulated mini flash crash are very straightforward:

\begin{itemize}
\item At 12:08:38, a spiking trader is activated to bring downward shocks to the market. The spiking trader submits sell orders to the market in the next 2 seconds. Those sell pressure in the 2-second interval is mainly absorbed by market makers.
\item Having absorbed the sell orders from spiking trader, one market maker accumulates a relatively large inventory at 12:08:40 and the inventory limit is reached. For risk management purposes, the market maker decides to reduce the position and then temporarily withdraw from the market. The sell orders from this market maker are digested by other market makers.
\item The same high-frequency dynamic happens between 12:08:40 and 12:08:42. During this small time period, the same positions transfer between different market makers, creating the "hot-potato" effect on a smaller scale. Several market makers withdraw from the market temporarily after emptying their inventory.
\item At 12:08:42, only 4 seconds after the spiking trader is activated, the market suffers from the liquidity loss because of the withdrawal of some market makers. The sell orders from the remaining market makers create dramatic market impacts due to thin liquidity, dragging the price down for more than 60 basis points in 5 seconds. 
\item At 12:08:47, the price has dropped 80 basis points compared to the price level at 12:08:38. The price then stops dropping and then gradually bounces back to the original level.
\end{itemize}

\hspace{0.3cm} Note that in the above simulation, the fundamental traders are configured to have a trading interval of 100 steps, which is the calibrated value from Section~\ref{modelcalibration}. The specific simulated mini flash crash event happens in a time horizon that is shorter than the horizon over which lower frequency fundamental traders observe the market. This is shown in Figure~\ref{minidetailed}, where the inventory for fundamental traders hardly changed during the mini flash crash event. The analysis shows that the mini flash crash events result from the procyclical behaviours among high-frequency market makers, precipitated by price shocks that are created by spiking traders or other market participants.

\begin{figure}[H]
\centering
\includegraphics[width=15cm, height=4cm, angle=0]{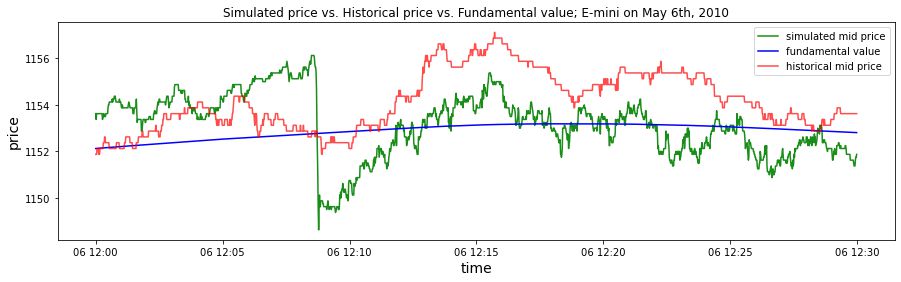}
\caption{One typical simulated mini flash crash event. Red line is the historical price; green line is the simulated price; blue line is the fundamental value}
\label{miniflashcrash}
\end{figure}

\subsection{Conditions for Mini Flash Crash Scenarios}
\hspace{0.3cm} In Section~\ref{conditionforlargeflashcrash} we explore how different conditions influence the amplitude of large flash crash events. Here we address the question that how those conditions influence the characteristics of mini flash crash events. Since the institutional trader is removed from the simulation, two conditions that are unrelated to the institutional trader are considered: market makers inventory limit and trading frequency of fundamental traders. Unlike the colossal flash crashes that happen extremely rarely, mini flash crashes occur more frequently in modern financial markets where high-frequency trading contributes to a vast portion of transactions. Consequently, the characteristic of mini flash crash events is twofold: the frequency for the occurrence of mini flash crash events and the crash severity once a mini flash crash occurs.

\hspace{0.3cm} To explore the characteristics of mini flash crashes under different conditions, Monte Carlo simulation is carried out. For the purpose of consistency, the simulation also mimics the E-mini futures market on May 6th. The simulation spans from 8:00 to 12:30 on May 6th, 2010, excluding the afternoon trading session to reduce the influence of the large historical flash crash event. Except for the two model parameters that represent the two conditions that are of interest, other model parameters are assigned the same value as the calibration results and are strictly kept fixed\footnote{Due to no calibration results, the model parameters for the newly introduced spiking trader are given values heuristically.}. For each parameter combination, which represents a specific condition, we run 60 simulations and calculate the average frequency for the occurrence of mini flash crash events and the average amplitude for the mini flash crash events.

\hspace{0.3cm} The only remaining question to solve is how to count the mini flash crashes and how to measure the amplitude of a mini flash crash. Following \cite{karvik2018deeds}, a mini flash crash is classified as a $k$ standard deviation move in price, which reverses over a horizon that is less than certain time periods. The $k$ has value 2, 3, 4 in our experiments. Note that because the spiking trader can create both upward and downward price shock, we also consider the upward "flare up" event as another form of mini flash crash\footnote{The definition for "flare up" and "flash crash" are symmetrical: "flare up" is a rapid price rise followed by a price drop.}. Specifically, in our experiments a $k$-sd mini flash crash is defined to be the peak or trough price behaviour inside a 10-minute interval, where the price moves more than $k$ standard deviations and then bounces back. The $k$ standard deviation is calculated by reference to the minute-level return. Inspired by topography, the amplitude of the mini flash crash event is then defined to be the prominence of the peak or trough price trajectory. A specific method to calculate the prominence can be found in \cite{2020SciPy-NMeth}.

\begin{figure}[H]
\centering
\includegraphics[width=16cm, height=5.2cm, angle=0]{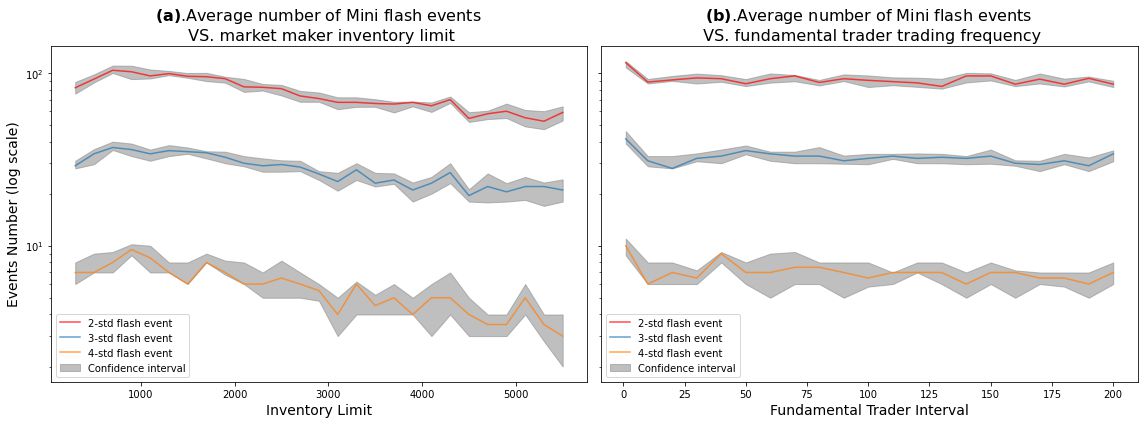}
\caption{\textbf{(a)}: Relationship between the average number of mini flash crashes per simulation and the inventory limit of market maker. \textbf{(b)}: Relationship between the average number of mini flash crashes per simulation and the trading frequency of fundamental traders. }
\label{miniconditions}
\end{figure}

\begin{figure}[H]
\centering
\includegraphics[width=16cm, height=5.2cm, angle=0]{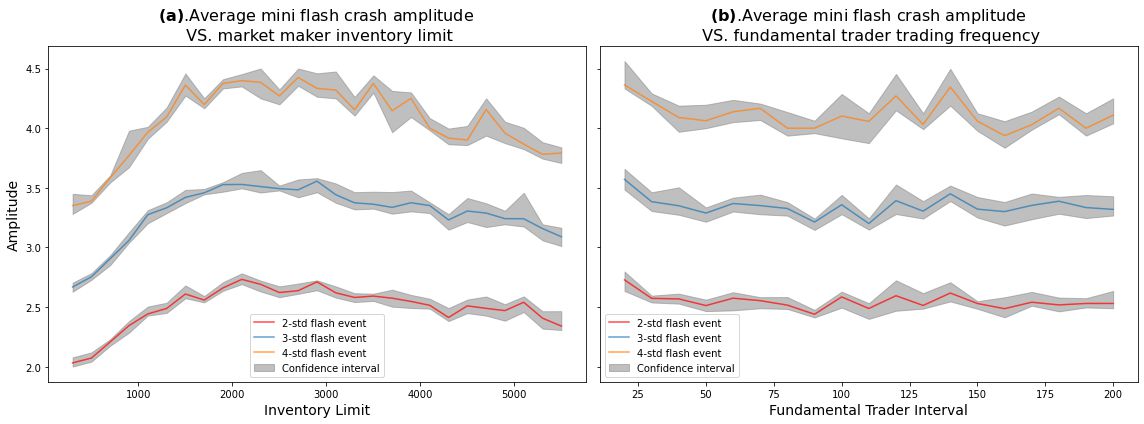}
\caption{\textbf{(a)}: Relationship between the average amplitude of simulated mini flash crashes and the inventory limit of market maker. \textbf{(b)}: Relationship between the average amplitude of simulated mini flash crashes and the trading frequency of fundamental traders. }
\label{miniampconditions}
\end{figure}

\hspace{0.3cm} We calculate the average number of the occurrence of mini flash crash and the average amplitude for the mini flash crash events under different conditions. The results are shown in Figure~\ref{miniconditions} and Figure~\ref{miniampconditions}.

\hspace{0.3cm} Figure~\ref{miniconditions} presents how the average number of mini flash crashes per simulation changes when the inventory limit of market makers and trading frequency of fundamental traders vary. As shown in panel (a), increasing the inventory limit will decrease the average number of mini flash crashes in our simulation, indicating a lower frequency for the occurrence of mini flash crash events. This result is consistent with our intuition. The larger the inventory limit for market makers, the more selling pressure market makers are able to absorb. Thus the "hot-potato" effect is less probable to appear with a larger inventory limit for market makers, resulting in fewer mini flash crashes. As for the trading frequency of fundamental traders, however, the frequency of the occurrence for mini flash crash events barely changes when we vary the trading frequency parameter of fundamental traders. This is presented in panel (b) of Figure~\ref{miniconditions}. One possible explanation is that mini flash crash events happen in an extremely short time scale and the amplitude is relatively small. Consequently, fundamental traders are hardly able to nor willing to participate during the course of the mini flash crash events. 

\hspace{0.3cm} Figure~\ref{miniampconditions} presents how the average amplitude of mini flash crashes changes when the inventory limit of market makers and trading frequency of fundamental traders vary. As demonstrated by Panel (a) in Figure~\ref{miniampconditions}, the functional relationship between mini flash crash amplitude and market maker inventory limit is not monotonous. The average mini flash crash amplitude is an increasing function with regard to market maker inventory limit when the inventory limit is small, while it turns into a decreasing function when the market maker inventory limit is large enough. The logic here is very similar to the analysis in Section~\ref{conditionforlargeflashcrash}. While the market maker inventory limit is small, increasing the inventory limit will result in more selling pressure when the inventory limit is hit, leading to a larger crash amplitude. However, a large enough market maker inventory limit would absorb most shocks in the market, restricting the scale of the mini flash crash events. Moving on to panel (b) in Figure~\ref{miniampconditions}, we observe no obvious influence of fundamental traders trading frequency on the amplitude of mini flash crash events. This phenomenon once again demonstrates that fundamental traders hardly participate during the course of mini flash crash events.

\section{Conclusion and Future Work} \label{conclusion}
\subsection{Summary of Achievements}
\hspace{0.3cm} A novel high-frequency agent-based financial market simulator is implemented to generate a realistic high-frequency simulated financial market. Each simulation step corresponds to 100 milliseconds in the real-world trading environment. Full exchange protocols (limit order books) are implemented to simulate the order matching process. In this way, we provide a microstructure model of a single security traded on a central limit order book in which market participants follow fixed behavioural rules. The model is calibrated using the machine learning surrogate modelling approach. Statistical test and moment coverage ratio results show that the simulation is capable of reproducing realistic stylised facts in financial markets.
 
\hspace{0.3cm} The simulator is then employed to explore the dynamics during flash crash episodes and the conditions that affect flash crash episodes. Under the framework of the proposed high-frequency agent-based financial market simulator, the 2010 Flash Crash is realistically simulated by introducing an institutional trader that mimics the real-world Sell Algorithm on May 6th, 2010. We investigate the market dynamics during the simulated flash crash and show that the simulated dynamics are consistent with what happened in historical flash crash scenarios. We then explore the conditions that could influence the characteristics of the 2010 Flash Crash. It is found that three conditions significantly affect the amplitude of the 2010 Flash Crash: the percentage of volume of the Sell Algorithm, market maker inventory limit, and the trading frequency of fundamental traders. In particular, we found that the relationship between the amplitude of the simulated 2010 Flash Crash and the POV of the Sell Algorithm is not monotonous, and so is the relationship between the amplitude and the market maker inventory limit. For the trading frequency of fundamental traders, the higher the frequency, the smaller the amplitude of the simulated 2010 Flash Crash.
 
\hspace{0.3cm} Similar analysis is carried out for mini flash crash events. An innovative type of trader called "Spiking Trader" is introduced to the agent-based financial market simulator, creating more price shock and precipitating more mini flash crash events. Market dynamics for a typical simulated mini flash crash event are analysed. We also explore the conditions that could influence the characteristics of mini flash crash events. Experimental results show that market maker inventory limit significantly affects both the frequency and amplitude of mini flash crash events. However, the trading frequency of fundamental traders shows no obvious influence on mini flash crash events in our experiments.

\subsection{Future Work}

\hspace{0.3cm} During the 2010 Flash Crash, there are lots of cross-market arbitrageurs who transferred the selling pressure to the equities markets by opportunistically buying E-Mini contracts and simultaneously selling products like SPY. This cross-market mechanism has not yet been implemented in our agent-based modelling framework. One future direction is to implement simulated markets for two correlated securities and explore the contagion during stressed scenarios. Another extension is to use the proposed agent-based financial market simulation framework for examining how regulatory policy interventions could influence the current market dynamics. For example, whether a circuit breaker in the market would help stabilize financial markets and curb the severity of flash crash scenarios. Last but not the least, an examination of possible indicators of an imminent flash crash event is another interesting future extension of this work.




\bibliographystyle{myapacite}
\bibliography{references}

\newpage
\begin{appendices}
\section{Descriptions for All Model Parameters} \label{modelparameterdescription}

\begin{table}[h]
\centering
\caption{Descriptions for All Parameters involved in the proposed agent-based model}
\begin{tabular}{ | m{1.6cm}<\centering | m{5.7cm}<\centering | m{1.6cm}<\centering | m{5.7cm}<\centering |} 
\hline
Parameter & Description & Parameter & Description\\
\hline
$\kappa_1$ & FT: Coefficient for linear demand component & $\kappa_2$ & FT: Coefficient for polynomial demand component \\ 
\hline
$N_{FT}$ & FT: Number of FT  & $S^{FT}_{interval}$ & FT: Interval between fundamental trading activities. \\
\hline
$\beta_L$ & LMT: Coefficient for demand calculation  & $\alpha_L$ & LMT: Coefficient for demand calculation\\
\hline
$N_{LMT}$ & LMT: Number of LMT  & $\beta_S$ & SMT: Coefficient for demand calculation \\ 
\hline
 $N_{SMT}$ & SMT: Number of SMT  &  $\alpha_S$ & SMT: Coefficient for demand calculation\\
\hline
$\sigma_{NT}$ & NT: Coefficient for demand calculation & $N_{NT}$ & NT: Number of NT  \\ 
\hline
$\gamma$ & LMT \& SMT: Coefficient for demand calculation & $\delta$ & NT, LMT \& SMT: Limit order cancellation rate\\ 
\hline
$\rho$ & NT, LMT \& SMT: Ratio between probability of submitting a market order and probability of submitting a limit order & $V$ & NT, FT, LMT, SMT, MM: Order volume \\
\hline
$\mu_{\ell}$ & NT, LMT \& SMT: Mean of the log-normal distribution from which limit order price distance is sampled & $\Sigma_{\ell}$ & NT, LMT \& SMT: Standard deviation of the log-normal distribution from which limit order price distance is sampled \\ 
\hline
$N_{MM}$ & MM: Number of MM  & $\delta_{MM}$ & MM: Limit order cancellation rate \\ 
\hline
$\theta_{MM}$ & MM: Probability of submitting a quote & $p^{MM}_{edge}$ & MM: Maximum price distance of submitted order  \\ 
\hline
$\varepsilon^{MM}_{limit}$ & MM: Position limit & $\varepsilon^{MM}_{safe}$ & MM: Safe position level \\ 
\hline
$\varepsilon^{MM}_{rest}$ & MM: Time length for the trading suspension & $N_{ST}$ & ST: Number of ST \\ 
\hline
$N_{spike}$ &ST: Number of consecutive orders to be submitted after ST is activated & $\mu_{spike}$ & ST: Probability of being activated in a certain simulation step \\ 
\hline
$V_{spike}$ & ST: Order volume & $r$ & INS: Target execution rate\\ 
\hline
$Q$ & INS: Number of initial inventory & $n$ & INS: Number of seconds between consecutive market sell orders \\ 
\hline
\end{tabular}
\end{table}

\newpage
\section{Values for Fixed Model Parameters in Calibration} \label{fixedparameters}

\begin{table}[h]
\centering
\caption{Values for fixed model parameters in calibration}
\begin{tabular}{ | m{2.5cm}<\centering | m{2.5cm}<\centering | m{2.5cm}<\centering | m{2.5cm}<\centering | m{2.5cm}<\centering |} 
\Xhline{2pt}
Parameter & $\kappa_1$ & $\kappa_2$ & $N_{FT}$ & $S^{FT}_{interval}$\\
\hline
Value & N.A. & N.A. & 30 & 100\\
\Xhline{2pt}
Parameter & $\beta_L$ & $\alpha_L$ & $N_{LMT}$ & $\beta_S$\\
\hline
Value & N.A. & 0.001 & 30 & N.A.\\
\Xhline{2pt}
Parameter & $N_{SMT}$ & $\alpha_S$ & $\sigma_{NT}$ & $N_{NT}$\\
\hline
Value & 30 & 0.9 & N.A. & 30\\
\Xhline{2pt}
Parameter & $\gamma$ & $\delta$ & $\rho$ & $V$\\
\hline
Value & 10 & 0.005 & 0.2 & 100\\
\Xhline{2pt}
Parameter & $\mu_{\ell}$ & $\Sigma_{\ell}$ & $N_{MM}$ & $\delta_{MM}$\\
\hline
Value & N.A. & 0.3  & 20 & 0.05\\
\Xhline{2pt}
Parameter & $\theta_{MM}$ & $p^{MM}_{edge}$ & $\varepsilon^{MM}_{limit}$ & $\varepsilon^{MM}_{safe}$\\
\hline
Value & N.A. & 4 & 5000 & 101\\
\Xhline{2pt}
Parameter & $\varepsilon^{MM}_{rest}$ & $N_{ST}$ & $N_{spike}$ & $\mu_{spike}$\\
\hline
Value & 12000 & N.A. & N.A. & N.A.\\
\Xhline{2pt}
Parameter & $V_{spike}$ & $r$ & $Q$ & $n$ \\
\hline
Value & N.A. & N.A. & N.A. & N.A. \\
\Xhline{2pt}
\end{tabular}
\end{table}

\section{Values for Model Parameters in 2010 Flash Crash Simulation} \label{may6thparameterapp}

\begin{table}[h]
\centering
\caption{Values for model parameters in 2010 Flash Crash simulation}
\begin{tabular}{ | m{2.5cm}<\centering | m{2.5cm}<\centering | m{2.5cm}<\centering | m{2.5cm}<\centering | m{2.5cm}<\centering |} 
\Xhline{2pt}
Parameter & $\kappa_1$ & $\kappa_2$ & $N_{FT}$ & $S^{FT}_{interval}$\\
\hline
Value & 0.1390 & 0.4562 & 30 & 100\\
\Xhline{2pt}
Parameter & $\beta_L$ & $\alpha_L$ & $N_{LMT}$ & $\beta_S$\\
\hline
Value & 0.3017 & 0.001 & 30 & 0.1273\\
\Xhline{2pt}
Parameter & $N_{SMT}$ & $\alpha_S$ & $\sigma_{NT}$ & $N_{NT}$\\
\hline
Value & 30 & 0.9 & 0.3403 & 30\\
\Xhline{2pt}
Parameter & $\gamma$ & $\delta$ & $\rho$ & $V$\\
\hline
Value & 10 & 0.005 & 0.2 & 100\\
\Xhline{2pt}
Parameter & $\mu_{\ell}$ & $\Sigma_{\ell}$ & $N_{MM}$ & $\delta_{MM}$\\
\hline
Value & 1.9349 & 0.3  & 20 & 0.05\\
\Xhline{2pt}
Parameter & $\theta_{MM}$ & $p^{MM}_{edge}$ & $\varepsilon^{MM}_{limit}$ & $\varepsilon^{MM}_{safe}$\\
\hline
Value & 0.6624 & 4 & 7000 & 101\\
\Xhline{2pt}
Parameter & $\varepsilon^{MM}_{rest}$ & $N_{ST}$ & $N_{spike}$ & $\mu_{spike}$\\
\hline
Value & 12000 & N.A. & N.A. & N.A.\\
\Xhline{2pt}
Parameter & $V_{spike}$ & $r$ & $Q$ & $n$ \\
\hline
Value & N.A. & 9\% & 120,000 & 12 \\
\Xhline{2pt}
\end{tabular}
\end{table}

\newpage

\section{Values for Model Parameters in Mini Flash Crash Simulation} \label{miniparameterapp}

\begin{table}[h]
\centering
\caption{Values for model parameters in Mini Flash Crash simulation}
\begin{tabular}{ | m{2.5cm}<\centering | m{2.5cm}<\centering | m{2.5cm}<\centering | m{2.5cm}<\centering | m{2.5cm}<\centering |} 
\Xhline{2pt}
Parameter & $\kappa_1$ & $\kappa_2$ & $N_{FT}$ & $S^{FT}_{interval}$\\
\hline
Value & 0.1390 & 0.4562 & 30 & 100\\
\Xhline{2pt}
Parameter & $\beta_L$ & $\alpha_L$ & $N_{LMT}$ & $\beta_S$\\
\hline
Value & 0.3017 & 0.001 & 30 & 0.1273\\
\Xhline{2pt}
Parameter & $N_{SMT}$ & $\alpha_S$ & $\sigma_{NT}$ & $N_{NT}$\\
\hline
Value & 30 & 0.9 & 0.3403 & 30\\
\Xhline{2pt}
Parameter & $\gamma$ & $\delta$ & $\rho$ & $V$\\
\hline
Value & 10 & 0.005 & 0.2 & 100\\
\Xhline{2pt}
Parameter & $\mu_{\ell}$ & $\Sigma_{\ell}$ & $N_{MM}$ & $\delta_{MM}$\\
\hline
Value & 1.9349 & 0.3  & 5 & 0.05\\
\Xhline{2pt}
Parameter & $\theta_{MM}$ & $p^{MM}_{edge}$ & $\varepsilon^{MM}_{limit}$ & $\varepsilon^{MM}_{safe}$\\
\hline
Value & 0.6624 & 4 & 4000 & 101\\
\Xhline{2pt}
Parameter & $\varepsilon^{MM}_{rest}$ & $N_{ST}$ & $N_{spike}$ & $\mu_{spike}$\\
\hline
Value & 12000 & 2 & 4 & 0.005\\
\Xhline{2pt}
Parameter & $V_{spike}$ & $r$ & $Q$ & $n$ \\
\hline
Value & 100 & N.A. & N.A. & N.A. \\
\Xhline{2pt}
\end{tabular}
\end{table}

\end{appendices}

\end{document}